\documentclass[journal,12pt,onecolumn,draftclsnofoot]{IEEEtran}
\IEEEoverridecommandlockouts   
\usepackage{algorithm,algorithmic,subfigure,amsfonts,amsmath,amssymb,mathrsfs,amsthm,graphicx,color,tikz,multirow,url,footmisc,epstopdf}
\usepackage[normalem]{ulem} 
\usepackage[utf8]{inputenc}
\usepackage{bbding}
\usepackage{mathtools}
\usepackage{algorithm}
\usepackage{algorithmic}
\usepackage{soul}
\usepackage{caption}
\usepackage{siunitx}
\usepackage{eurosym}
\usepackage{verbatim}
\usepackage{tikz,fp}
\usepackage{cancel}
\usepackage{enumitem}
\usepackage{amsmath,amssymb,amstext}

\usepackage{booktabs}
\usepackage{soul}

\usepackage{framed}
\usepackage{eurosym}
\usepackage{amsmath}
\usepackage{footnote}
\usepackage{cite}
\usepackage{pgfplots}
    \usetikzlibrary{
        matrix,
    }
    \pgfplotsset{
        compat=1.3,
    }
\usetikzlibrary{calc,patterns,angles,quotes}
\usetikzlibrary{shapes}
\usetikzlibrary{plotmarks}

\usetikzlibrary{shadows,patterns,shapes,backgrounds,automata}
\usetikzlibrary{fit,decorations.pathmorphing,plotmarks} 
\tikzstyle{every picture}+=[remember picture] 

\graphicspath{{./Figures/}}

\newcommand{\specialcell}[2][c]{\begin{tabular}[#1]{@{}c@{}}#2\end{tabular}}

\newtheorem{lemma}{Lemma}
\newtheorem{remark}{Remark}
\newtheorem{definition}{Definitions}

\usepackage{tikz}
\usetikzlibrary{shadows,arrows}
\pgfdeclarelayer{background}
\pgfdeclarelayer{foreground}
\pgfsetlayers{background,main,foreground}
 
\tikzstyle{materia}=[draw, fill=blue!20, text width=6.0em, text centered,
  minimum height=1.5em,drop shadow]
\tikzstyle{practica} = [materia, text width=8em, minimum width=10em,
  minimum height=3em, rounded corners, drop shadow]
\tikzstyle{texto} = [above, text width=6em, text centered]
\tikzstyle{linepart} = [draw, thick, color=black!50, -latex', dashed]
\tikzstyle{line} = [draw, thick, color=black!50, -latex']
\tikzstyle{ur}=[draw, text centered, minimum height=0.01em]
 
\usepackage{ellipsis}
\usetikzlibrary{calc}
\usetikzlibrary{decorations.pathreplacing,decorations.markings,shapes.geometric}
\tikzset{naming/.style={align=center,font=\small}}
\tikzset{antenna/.style={insert path={-- coordinate (ant#1) ++(0,0.25) -- +(135:0.25) + (0,0) -- +(45:0.25)}}}
\tikzset{station/.style={naming,draw,shape=dart,shape border rotate=90, minimum width=10mm, minimum height=10mm,outer sep=0pt,inner sep=3pt}}
\tikzset{mobile/.style={naming,draw,shape=rectangle,minimum width=12mm,minimum height=6mm, outer sep=0pt,inner sep=3pt}}
\tikzset{radiation/.style={{decorate,decoration={expanding waves,angle=90,segment length=4pt}}}}

\newcommand{\MBS}[1]{%
\begin{tikzpicture}
\node[station] (base) {#1};

\draw[line join=bevel] (base.100) -- (base.80) -- (base.110) -- (base.70) -- (base.north west) -- (base.north east);
\draw[line join=bevel] (base.100) -- (base.70) (base.110) -- (base.north east);

\draw[line cap=rect] ([xshift=-.1768cm,yshift=.6pt]base.north -| base.right tail) [antenna=1];
\draw[line cap=rect] ([yshift=.6pt]ant1 |- base.north) -- node[above,shape=rectangle,inner ysep=+.3333em]{\dots} ([xshift=.1768cm,yshift=.6pt]base.north -| base.left tail) [antenna=2];

\end{tikzpicture}
}

\newdimen\bpt
\def\mobile#1{\leavevmode 
   \bpt=#1bp \hbox to7\bpt{\kern1\bpt \lower1\bpt\vbox to12\bpt{}%
      \pdfliteral{q #1 0 0 #1 0 0 cm 1 j 2 w 0 0 5 10 re B 
         1 g 1 G  1 w .3 1.8 4.4 7 re B 
         1.5 w 2.5 .2 0 .1 re B .3 w 1.7 10 1.6 0 re B Q}%
      \hss}}

\pgfdeclaredecoration{lightning bolt}{draw}{
\state{draw}[width=\pgfdecoratedpathlength]{
  \pgfpathmoveto{\pgfpointorigin}%
  \pgfpathlineto{\pgfpoint{\pgfdecoratedpathlength*0.6}%
    {-\pgfdecoratedpathlength*.1}}%
  \pgfpathlineto{\pgfpoint{\pgfdecoratedpathlength*0.55}{0pt}}%
  \pgfpathlineto{\pgfpoint{\pgfdecoratedpathlength}{0pt}}%
  \pgfpathlineto{\pgfpoint{\pgfdecoratedpathlength*0.4}%
    {\pgfdecoratedpathlength*.1}}%
  \pgfpathlineto{\pgfpoint{\pgfdecoratedpathlength*0.45}{0pt}}%
  \pgfpathclose%
}%
}

\title{A General Framework for Airplane Air-to-Ground Communications in mmWave and Microwave Bands}
\author{Ararat Shaverdian, Shahram Shahsavari, and Catherine Rosenberg,~\IEEEmembership{Fellow,~IEEE} 

\thanks{This is a longer version of a paper entitled ``Air-to-Ground Cellular Communications For Airplane Maintenance Data Offloading'' that is under submission.}
\thanks{A. Shaverdian, S. Shahsavari, and C. Rosenberg are with the Department of Electrical and Computer Engineering, University of Waterloo, Waterloo, On, Canada. Email: \{ashaverd, shahram.shahsavari, cath\}@uwaterloo.ca}
}

\begin{document}

\maketitle 
 

\begin{abstract}
Airplane sensors and on-board equipment collect an increasingly large amount of maintenance data during flights that are used for airplane maintenance. We propose to download part of the data during airplane's descent via a cellular base station (BS) located at the airport.  
We formulate and solve an offline optimization problem to quantify how much data can be offloaded in a {\it non-dedicated band} while ensuring that the interference at the terrestrial BSs in the vicinity of the airport remains below a maximum allowable threshold. Our problem allows for adaptive tuning of transmit power, number of frequency channels to be used, and beamforming according to the position of the plane on the descent path. 
Our results show that during the last 5 minutes of descent, in the microwave band the plane can offload up to 5 GB of maintenance data in a 20~MHz band, while in the mmWave band the plane can offload up to  24 times more data in a 1~GHz band. Beamforming, power and bandwidth tuning are all crucial in maintaining a good performance in the mmWave band while in the microwave band, dynamic tuning of bandwidth does not improve the performance much. 
\end{abstract}


\section{Introduction} 
With the modernization of aviation systems, a surge in on-board maintenance data is expected~\cite{1:tb,1:tb:2, 1:TB:thesis}. This data is collected by on-board sensors to help airliners better monitor and maintain their airplanes. Today maintenance data is downloaded from on-board memory by airliner crew through a cable while the airplane is on the ground at the gate and then transferred to the airliner's data center for processing. However, with the foreseen increase in the maintenance data volume estimated at 1 terabyte per plane~\cite{1:tb,1:tb:2}, it is unlikely that this way of offloading data will remain the airliners' preferred choice. This is because it can potentially delay the plane's take-off time, and reduce the revenue which is directly proportional to airborne time. 
Air-to-ground communications (A2G) can provide a fast and inexpensive  solution to relaying part of the data from the airplane to the airliner's data centre by enabling the airliner  to start downloading the maintenance data prior to touchdown while the airplane is still airborne. 
In this paper, we propose to download part of the data during the descent phase, which corresponds to the time between the end of cruising phase and  start of touchdown, using a designated terrestrial cellular base station (BS) at the airport. This is a good time for downloading the data since from an airliner's perspective, by the end of the cruising phase, plane sensors would have collected most of the in-flight measurements for maintenance. The remainder of the data can then be offloaded during taxiing or, at the gate, by the airliner crew.

In this paper, 
our goal is to quantify how much data can be offloaded during the descent phase under the assumption that the operator uses a non-dedicated spectrum in either microwave or mmWave bands. We are not concerned with online processes, such as scheduling, and the complexities involved in the operation of the network. In that sense, this is an offline study suitable for planning phase.

A non-dedicated band for A2G
implies that the spectrum is reused for terrestrial cellular users {\it in the uplink}. This assumption is both more conservative and realistic than a dedicated band (since spectrum is an expensive resource especially in microwave bands), although it implies that A2G over the shared band would create interference at non-A2G terrestrial BSs (TBSs), i.e., the BSs in the vicinity of the airport that are not used for A2G. A2G being a new and peripheral service has to co-exist with TBSs without causing too much interference. What we propose the operator do is limit the impact of A2G interference on its TBSs on the shared band by enforcing a maximum allowable interference power (coming from the plane) at those BSs. Then, given the two options of a non-dedicated spectrum in either mmWave or microwave band, the criteria for evaluating the system performance is not only efficiency but also the impact that A2G would have on TBSs. 




We consider a runway at an airport and a corresponding descent path. A network operator places one designated A2G BS (ABS) close to the middle of the runway at the airport to facilitate offloading the maintenance data of an airliner's plane during its descent using a cellular technology such as LTE-A 4G or 5G. A single plane enters the descent path and starts transmission $T_s$ seconds before landing until touchdown where it ceases A2G transmission. 
The operator has access to one of the following frequency bands that are shared by all of its BSs in the vicinity of the airport, including both the ABS and TBSs: a microwave sub-6 GHz carrier with a relatively small bandwidth or a mmWave carrier with a large bandwidth. 
We wish to solve the following research problem: {\it Given a carrier frequency and an associated bandwidth --- made up of $N_\text{sub}$ OFDMA frequency subchannels --- that is shared by the BSs of a cellular operator's network, the location of each BS in the operator's network, the type of antenna(s) installed at the BSs and the descending plane, a transmit power budget at the plane, a maximum allowable interference power threshold (from the plane) at the operator's TBSs, find the maximum amount of maintenance data that can be offloaded from the plane during its transmission time $T_s$ while ensuring that the interference at TBSs remains below the maximum allowable interference threshold.} In solving this problem, we consider the following generic types of antenna installed either at the plane or the ABS: a single omnidirectional or directional antenna, or a uniform planar array (UPA). Directional antennas are simple and characterized by a beamwidth and tilting angle which are typically quasi-static. UPAs, on the other hand, can adaptively create beam patterns that focus the radiated transmit power towards a desired direction and away from specific locations. 

We identify three parameters that the operator can tune to increase the amount of offloaded data while limiting the A2G interference: \textit{i}) {\it transmit power to use at a given time at the plane}, \textit{ii}) {\it number of subchannels to use at a given time}, and \textit{iii}) {\it beam pattern at the ABS and descending plane}. As the plane descends along the descent path, its position relative to each BS in the network changes and, so does the A2G interference. Fortunately, as the location of the BSs as well as the trajectory of the plane are known a priori, the operator can compute the plane's distance from each of these BSs, evaluate the channel conditions (e.g., path loss), and incorporate that information into deciding how much power, how many subchannels, and what beam pattern to use at each time instant during descent. 
What we propose the operator do is adaptively adjust the three parameters throughout the descent phase so that the A2G interference remains below the threshold while ensuring that the volume of downloaded data is maximized. If simple antennas are used at the plane and the ABS, the only parameters that can be adjusted are the transmit power and number of subchannels. 
If, however, the plane is equipped with a UPA there would be another degree of freedom - beside transmit power and subchannels - to control the A2G interference with, namely beamforming. In particular, the operator would be able to adaptively steer the radiated power of the plane away from the TBSs and toward the ABS through a narrow beam. To further enhance the signal-to-noise-ratio (SNR), UPA may also be used at the ABS.

We summarize our contributions in the following. 
We formulate and solve an offline optimization problem to compute the maximum data volume that can be offloaded from the plane at every time slot\footnote{As customary in cellular network, time is slotted and transmissions occur on a per time slot basis.} while the plane is on the descent without violating the A2G interference constraints, in any of the microwave and mmWave bands. 
Our formulated problem contains two original ideas. The first is that it enforces the control of the impact of A2G interference on legacy networks. The second is that it allows for adaptive tuning of transmit power, number of subchannels, and beamforming according to the position of the plane on the descent path. The two ideas are complementary in that the first one tries to protect the operator's TBSs from the airplanes transmissions at the cost of a possible drop in the offloaded data and the second one tries to minimize this drop at every time slot. Our formulated problem is non-convex and, hence, hard to solve.
We develop a solution technique to find an upper-bound and a feasible solution to the problem and show that they are close to each other. Contrarily to other problems where extracting a good feasible solution was easy (e.g., in~\cite{multicast-beamforming} and in~\cite{papc2018}), we had to develop an intricate method for obtaining good feasible solutions.  
 
The key messages from our results are the following. 
\begin{itemize}[leftmargin=*]
    \item \textbf{In the microwave band}, the operator can offload around 5~GB of  maintenance data during the last 5~minutes of descent on a 20-MHz band using the set of modulation and coding schemes (MCSs) of LTE-A. This value is very close to the system capacity\footnote{What we mean by system capacity is a data volume that cannot be exceeded even if there are no constraints on transmit power budget and maximum allowable interference power at TBSs.} for this set of MCSs. The bottleneck in the system capacity is the maximum data rate achievable with LTE-A and, if the operator wishes to increase the capacity in this band, the focus should be on using a technology that can support higher order MCSs. We use Shannon's formula to illustrate that a much higher gain  could be achieved by using higher-order MCSs. 
    Our results also indicate that it is best to use a large UPA at the BS and a small UPA at the plane. 
    We do not recommend using directional antennas on both ends since then the system performance deteriorates and becomes highly sensitive to the interference threshold. Furthermore, it is almost always optimal to use the entire bandwidth in the microwave band and so tuning the bandwidth does not improve the performance. 
    \item \textbf{In the mmWave band} (of 1-GHz), the operator can offload by using the LTE-A set of MCSs, up to 120~GB of  maintenance data during the last 5~minutes of descent which is 24 times larger than that in the microwave band (on a bandwidth which is 50 times larger).
    Because of the reduced coverage in the mmWave band, even a conservative interference threshold value does not impact the performance much, indicating that purchasing an exclusive band for A2G in the mmWave band would not improve the performance a lot. 
    However, as much a transmit power budget and as large a UPA as possible at both the ABS and the plane should be used to make up for this reduced coverage. Our results indicate that, because of the small coverage and large bandwidth on which the transmit power is spread, beamforming, power and bandwidth tuning are all crucial in this band. We also illustrate using Shannon's formula that higher-order MCSs would have a much more limited impact on the performance than in the mmWave band.
    
    \item \textbf{The choice of the antenna type} both at the plane and the ABS has a significant impact on the performance. For instance in a microwave band, having a sufficiently large antenna array (e.g., 25 antennas) on the plane provides a convenient control on the A2G interference by using beamforming appropriately. In such a case, our results reveal that providing an exclusive spectrum for A2G does not lead to considerable performance gains, whereas having a dedicated spectrum would significantly improve the performance if directional or omnidirectional antennas are used at the plane. Additionally, in mmWave bands, using large antenna arrays at both the plane and the ABS is necessary to compensate for the impact of path loss.
\end{itemize}




Notations: Throughout this paper, $w$ (or $W$), $\textbf{w}$, $\textbf{W}$, $(.)^H$, and $\textrm{Tr}(.)$ are used to denote a scalar $w$ (or $W$), vector $\textbf{w}$, matrix $\textbf{W}$, the complex conjugate transpose and trace operator, respectively.
$[\textbf{W}]_{m,n}$ denotes the element on row $m$ and column $n$ of matrix $\textbf{W}$. 
$\mathbb{R}^{n\times m}$, $\mathbb{C}^{n\times m}$ and $\mathbb{H}^n$ are used for the sets of $n$-by-$m$ dimensional real matrices, complex matrices and $n$-by-$n$ complex Hermitian matrices, respectively. 
$\textbf{W} \succeq 0$ indicates that $\textbf{W}$ is a positive semidefinite matrix. 
$\| . \|$ denotes the vector Euclidean norm. 
$\textbf{1}_\mathcal{A}(x)$ is the indicator function, i.e., $\textbf{1}_\mathcal{A}(x)=1$ if $x\in \mathcal{A}$ and $\textbf{1}_\mathcal{A}(x)=0$ if $x\notin \mathcal{A}$.
Finally, $\mathbf{I}_N$ is the unitary matrix with dimension $N$. 

\section{Related work}\label{literature}
There are a number of papers in the literature that study data delivery to on-board passengers, e.g.,~\cite{5g:a2g,cavdar,cavdar2} to cite a few. To the best of our knowledge, all of these papers focus is on the downlink\footnote{We refer to the transmissions from a BS to an aircraft as downlink and as uplink to the opposite direction.} as opposed to uplink which is the focus of our paper. There are a number of differences between maintenance data delivery in the uplink and passenger data in the downlink. In particular, for maintenance data delivery, it suffices to place one BS at the airport and offload part of the data during the descent. For passenger data delivery, however, BSs should be placed along the flight path to provide ubiquitous connectivity and, most definitely not near the airport, since users switch off their devices during descent and take-off. Furthermore, none of the papers consider the impact of A2G interference on the legacy terrestrial BSs. Our work is different in that it is the first comprehensive study that is tailored for maintenance data delivery and focuses on a general framework for interference management in both microwave and mmWave bands. 

There are numerous papers in the literature that study MIMO (Multiple Inputs Multiple Outputs) systems with UPA in cellular networks in the downlink (e.g.,~\cite{mimo-comp,zayani2019efficient,jin2019multicell,mimo-sec}). Most of these papers focus on multi-user MIMO where the BS, equipped with an antenna array, simultaneously serves multiple users, each equipped with a single and simple antenna. Our work inherently calls for a single-user MIMO in the uplink which is a departure from the conventional cellular systems. In particular, our system calls for an antenna array at the plane to adaptively direct the A2G power transmission away from TBSs and toward the ABS through beamforming. The ABS, in parallel, has to adaptively align its beam pattern with that of the plane.  In particular in~\cite{cavdar,cavdar2}, the authors study the impact of array size on performance in the context of A2G in the downlink for passenger traffic. The authors conclude that a large antenna array at the BSs --- transmitting to flying planes in the cruising altitude --- can improve the system throughput, and reduce beam-steering loss as well as inter-cell interference.
 


\section{System model}\label{sec:sys-model}
 
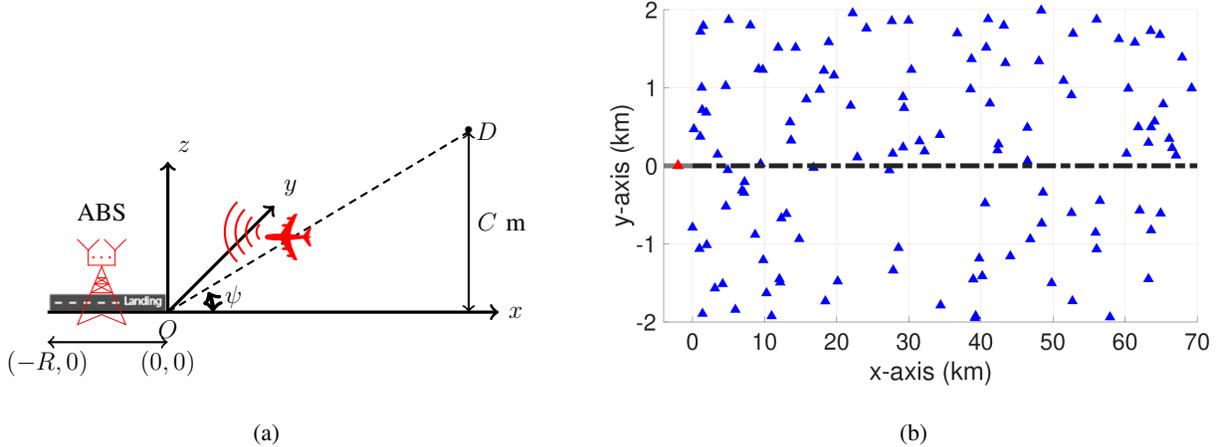
\begin{figure}[t!]
\subfigure[]{
    \begin{tikzpicture}[scale=0.8, every node/.style={scale=0.8}]
        \node [right] at ( 3,1.5 ) {$C$ m};    
        \draw[thick,<->] ( 3,0) -- ( 3,3 );
        \node[inner sep=0pt] (russell) at (-3,0.15)      
        {\includegraphics[width=0.12\textwidth]{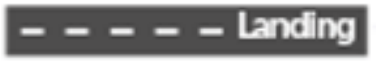}};
        \coordinate (origo) at (0,0);      
        \coordinate (land) at (-2,0);
        \coordinate (t) at (2,2);
        \draw[thick, densely dashed] (-2,0)  --  (3,3) node (path) {};
        \draw[very thick, ->] (-4,0) -- (3.5,0) node (x) [black,right] {$x$};
        \node[red] (plane) at (0,1.25) {\huge \reflectbox{\Plane}}; 
        \node [below] at (-2,0) {$O$};    
        \node [below] at (3.3,3.3) {$D$}; 
        \node [black] at (3,3) {\textbullet};
        \draw[thick,<->] ( -2,-0.5) -- ( -4,-0.5); 
        \node [below] at ( -2,-0.5) {$(0,0)$};
        \node [below] at ( -4,-0.5) {$(-R,0)$};       
        \pic [draw, ultra thick, <->, "$\psi$", angle eccentricity=1.5, angle radius=0.75cm] {angle = origo--land--t}; 
        \draw[very thick,->]   (-2,0) -- ++ (45:2.5) node[above right] {$y$};
        \draw[very thick,->]   (-2,0) -- ++ (90:2.5) node[above right] {$z$};
        \node[very thick,red,label=above:ABS] at (-3.1,0.5) {\MBS{}}; 
        \draw[thick,red,radiation,decoration={angle=45}] ([xshift=-.35cm,yshift=-.45cm]plane.north) -- +(180:0.75); 
        
    \end{tikzpicture}
    \label{fig:descent}
    }
    \subfigure[]{ 
        \includegraphics[scale=0.18]{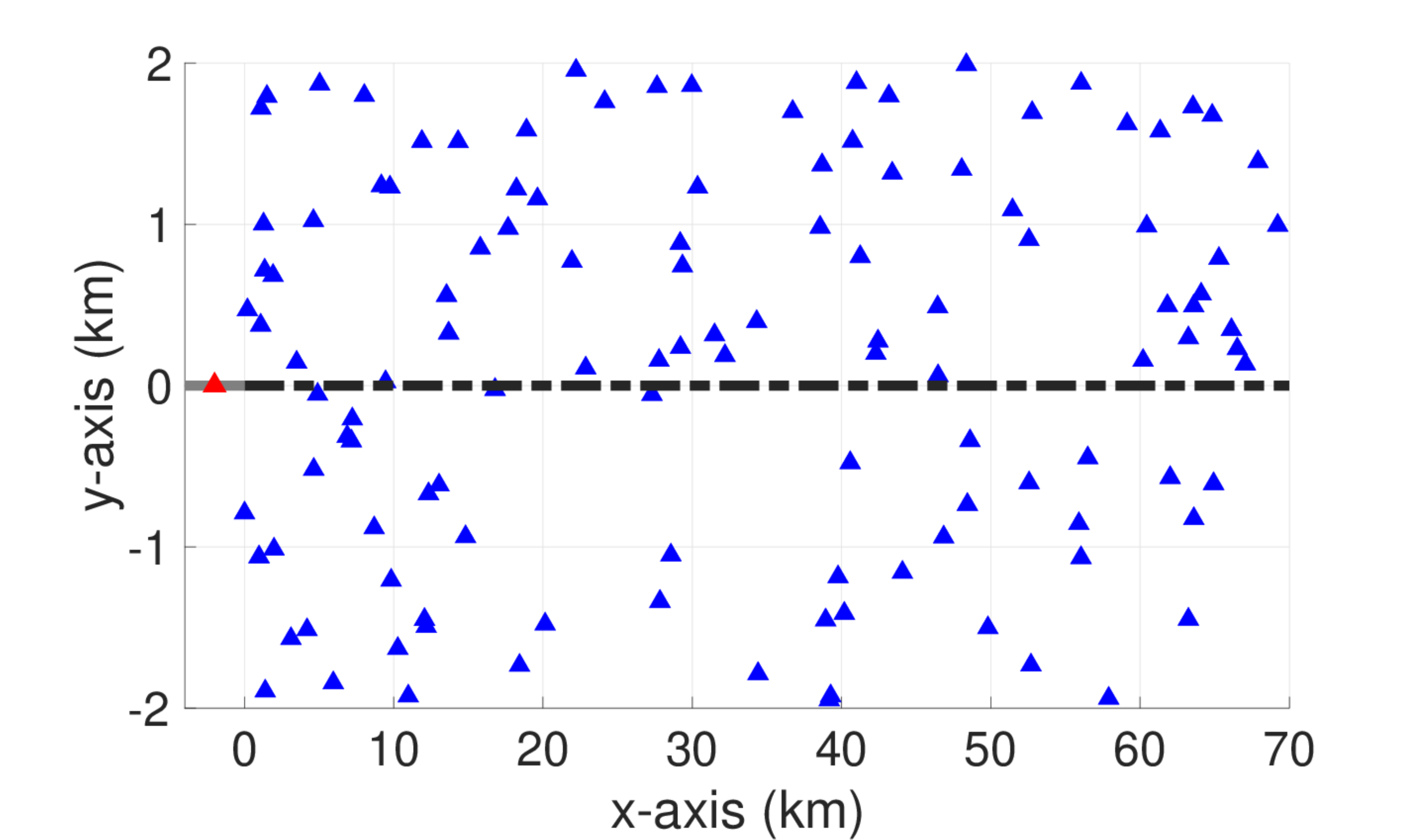} 
          \label{diag}} 
    \caption{An example of the system under study: (a) The dark dashed line denotes the active descent path with a pitch angle $\psi$. The plane transmits data to the designated A2G BS (ABS) located along the runway and will stop transmitting once it lands at the origin $O$. The data traffic is from the plane to the ABS. (b) The figure shows the bird's eye view of the system under study. The red triangle denotes the ABS at the airport. The blue ones are the interfered TBSs. The dashed line is the projection of the descent path on x-y plane. The point (0,0) is the touchdown point. The grey line is the runway. }
\end{figure}

We consider an airport with a runway of length $R$ as shown in Fig.~\ref{fig:descent}. Each end of the runway corresponds to the extreme point of a potential descent path which is the trajectory that a plane takes from the point where it leaves its cruising altitude until touchdown as illustrated in Fig.~\ref{fig:descent}. A plane approaching the airport can land from either end of the runaway but only one of these two descent paths is active at a time. The descent path and the descending approach are usually pre-determined and assigned to approaching planes by a mutual agreement between the pilot and control tower. Descending approach characterizes the mobility model (e.g., velocity profile) of the plane during its descent.  We consider a single plane on the descent path that starts transmitting as much maintenance data as possible before landing to a designated BS deployed by a cellular operator at the airport for this purpose, which we refer to as the ABS. 

We use the Cartesian coordinate system, with the fictitious point $O$ denoting the origin being located at the touchdown location, as depicted in Fig.~\ref{fig:descent}. Let $D$ be the point where the plane leaves its cruising altitude of $C$\footnote{All distances are in meters.} and starts descending while moving toward the airport to land. We define the descent path to be the line segment between the point $D$ and the origin $O$. Also, we define the runway to be the line segment between the origin $O$ and point $(-R,0)$. 
Lastly, we define the pitch angle of the descent path, denoted by $\psi$, to be the complement of the angle between the descent path and runway as illustrated in Fig.~\ref{fig:descent}. 
The network operator places the ABS in the middle of the runway at point $(-\frac{R}{2},0)$ in the line of sight (LoS) of the plane\footnote{The justification for placing the ABS in the middle of the runway is that the operator can that way serve any of the two potential active descent paths at the airport by either electronically or mechanically rotating the antenna(s) of the ABS so that it faces the active descent path.}. We denote the ABS with index 0.

The network operator that operates the ABS also operates its own terrestrial cellular network on the ground. This network includes $N_\text{TBS}$ TBSs indexed with $i\in \mathcal{I} \triangleq \{1,2,\ldots,N_\text{TBS}\}$. Fig.~\ref{diag} shows an example of the network under study with 120 TBSs distributed randomly in the vicinity of the airport.
Since spectrum is an expensive resource (especially in the microwave band), airliners may be reluctant to purchase a dedicated spectrum for A2G. As a result, we assume that the A2G transmissions (i.e., from the plane to the ABS) occur in the same time-frequency resources as the uplink (UL) transmissions in the cellular network of the operator (i.e., from terrestrial users to TBSs).
Consequently, there is an interference caused by A2G transmissions on the cellular UL reception at each TBS. 
The advantage of using cellular UL resources for A2G is that the interference occurs at the TBSs whose locations are fixed and known.
Below, we provide the definitions of A2G and interfering links which we will use throughout the paper.
\begin{definition} [\textbf{A2G and Interfering Links}]
The A2G link is defined as the propagation channel between the plane and ABS. The interfering link (IL) $i\in \mathcal{I}$ is defined as the propagation channel between the plane and TBS $i$. 
\end{definition} 
We assume that the A2G and interfering links are LoS which is a reasonable approximation in a practical A2G system where the plane is airborne during the A2G transmission. 
We note that the signal transmitted by the plane may also be received by the ABS and TBSs through reflections by the ground. However, we assume that these reflections are second-order compared to the LoS path in terms of the received power and neglect them in our model for the sake of theoretical tractability.
We assume that the plane descent path trajectory is known and given which is commonly the case in practice. Therefore, each interference channel can be approximated with a LoS channel whose characteristics are known since they only depend on the location of the interfered TBS and the plane. Clearly, using cellular downlink (DL) resources does not provide such an advantage as in that case the interference occurs at the cellular users (i.e., cellular DL receivers) which are usually mobile and this will complicate the computation of the interference.  


Following the convention in cellular networks such as LTE, we consider an OFDM time-frequency domain and assume that the plane transmits its data to the ABS on a per time slot basis. The bandwidth, denoted by $B$, is divided into $N_\text{sub}$ subchannels with frequency span of $b$. Furthermore, the time is divided into time slots of duration $\Delta t$. In LTE, $\Delta t$ is 1 ms while in 5G it could be smaller. Since in practice $\Delta t $ is very small and the plane is in the LoS of the BSs (ABS and TBSs), we can assume that the distance and channel gain between the plane and a BS during a slot are fixed and we set them to be the infimum of all the distances and supremum of all the channel gains in one slot respectively. At each time slot we assume that A2G transmission can occur over all or a fraction of available $N_\text{sub}$ subchannels. The number of subchannels used at time slot $t$ is denoted by $M(t)$ and is one of the control variables which will be optimized to increase the volume of the offloaded data over the A2G link.

It should be noted that our approach works for both time-division duplexing (TDD) and frequency-division duplexing (FDD). In the case of TDD, the plane will only transmit in the time slots designated for UL while using the entire spectrum. In the case of FDD, the plane only uses the spectrum allocated for UL transmission while transmitting at every time slot. Next, we describe the A2G scenarios which will be studied in this paper.

\subsection{A2G Scenarios} \label{subsec:scenarios}
In our study we use three types of antennas: (\textit{i}) a single antenna with omnidirectional (Omni) directivity pattern, (\textit{ii}) a single  antenna with directional (Dir) directivity pattern\footnote{\textcolor{black}{Directional antennas are commonly used in cellular networks for sectorization.  The operator may tilt the antennas to adjust its coverage or manage interference on/from neighbouring cells.}}, and (\textit{iii}) a UPA.
A directional antenna typically consists of multiple smaller omnidirectional or directional antenna elements that collectively create a beam pattern that is fixed and invariant in the long term. Furthermore, 
directional (or Omni) antennas can be tilted towards a desired direction to further enhance the transmission (or reception) gain at the plane (or ABS). The tilting is typically invariant in the long term. Directional antennas provide a good and simple solution for A2G management with a reasonably good gain (usually around 18~dBi). However, they are not a suitable candidate if the operator wishes to change the beamwidth and directionality of the antennas adaptively (e.g., to control the interference at the TBSs adaptively). 
UPAs --- also known as smart antennas --- are different from directional antennas in that they can adaptively create beam patterns that focus the radiated transmit power towards a desired direction in real time which is ideal for our A2G system. 
UPAs are made of small omnidirectional or directional antenna elements of small gain (e.g., 0~dBi for omnidirectional and 8~dBi for directional elements). 
They are also more compact than other types of smart antennas --- such as linear arrays --- which makes them a good candidate when using a large number of antenna elements. 
Typically the larger the number of antenna elements used in a UPA, the narrower the beam pattern and the better the directionality of the antenna array.  
In terms of practicality, it is best to place a small-sized antenna at the plane and a larger UPA at the ABS. 
In the UPA case, we assume that the phase and amplitude of the input signal of each antenna can be modified as in active phased arrays~\cite{sturdivant2018systems}. Note that  various other array architectures such as passive phased arrays --- where one can only change the phase of the antenna input signals --- can also be considered. They are not the focus of this paper due to space constraint and are left as an avenue for future research. 

In this paper, we consider the following four scenarios:
\begin{itemize}[leftmargin=*]
    \item \textbf{Scenario 1 (Dir/Omni -- Dir/Omni)}: The plane and ABS each have a single directional or Omni antenna. \textcolor{black}{In that case, the A2G link is Single Input Single Output (SISO).}
    
    \item \textbf{Scenario 2 (Dir/Omni -- UPA)}: The plane has a single directional or Omni antenna while the ABS is equipped with a UPA of Omni or directional antenna elements. \textcolor{black}{In that case, the A2G link is Single Input Multiple Outputs (SIMO).}
    
    \item \textbf{Scenario 3 (UPA -- Dir/Omni)}: The plane has a UPA of Omni or directional antenna elements while the ABS has a single directional or Omni antenna.  \textcolor{black}{In that case, the A2G link is Multiple Inputs Single Output (MISO).}
    
    \item \textbf{Scenario 4 (UPA -- UPA)}: The plane and the ABS are each equipped with a UPA of Omni or directional antenna elements. 
    In that case, the A2G link is Multiple Inputs Multiple Outputs (MIMO).
\end{itemize}
Note that the antenna type at each TBS depends on the type of the cellular network and we will consider various antenna types such as single directional and antenna arrays for TBSs in our numerical evaluations.

We consider a generic channel model which can represent all of the described scenarios above. \textcolor{black}{We focus on a subchannel $\ell$ of bandwidth $b$.} In this generic channel model, we assume that the plane and the ABS are equipped with antenna arrays consisting of $N_\text{P}$ and $N_\text{A}$ antennas, respectively. As a result, the A2G link is a MIMO channel (or SISO if $N_\text{A}=N_\text{P}=1$) which can be characterized by the channel matrix  \textcolor{black}{$\textbf{H}_0(t,\ell) \in \mathbb{C}^{N_\text{A}\times N_\text{P}}$} at time slot $t$. Furthermore, as highlighted before, we assume that the ground reflections are significantly weaker that the LoS path. Therefore, we use a three dimensional LoS MIMO channel model~\cite{upa0} where the element $(m,n)$ of matrix $\textbf{H}_0(t,\ell)$, i.e., the scalar channel gain between antenna element $m$ of the plane and antenna element $n$ of the ABS, is defined as
\begin{align}
    [\textbf{H}_0(t,\ell)]_{m,n} =  \sqrt{\beta_0(f_\ell,t) G^\text{P}_{0}(t)G^\text{A}(t)}~e^{-j \frac{2\pi}{\lambda(\ell)} d_{m,n}(t)} \label{eq:a2g-chan},
\end{align}
where $f_{\ell}$ is the central frequency of \textcolor{black}{the subchannel $\ell$, $\lambda(\ell)$ is the associated wavelength}, $\beta_0(f_{\ell},t)$ is the path loss over A2G link in frequency $f_{\ell}$ at time slot $t$, and $d_{m,n}(t)$ is the distance between antenna $m$ of the plane and antenna $n$ of the ABS at time slot $t$. Furthermore, $G^\text{P}_{0}(t)$ is the transmit directivity gain of the plane's antenna elements, in the direction toward the ABS at time slot $t$. Similarly, $G^\text{A}(t)$ is the receive directivity gain of the ABS antenna elements, in the direction from the plane at time slot $t$. \textcolor{black}{There are two important points to note about Equation ~\eqref{eq:a2g-chan}. First, while channel coefficients are generally frequency dependent ($\lambda(\ell)$} in~\eqref{eq:a2g-chan} \textcolor{black}{depends on the frequency $f_\ell$), we assume that the variation of $\textbf{H}_0(t,\ell)$ over frequency subchannels is negligible. Hence, we replace $f_\ell$ with $f_c$ the central frequency of the whole band and remove the index $\ell$ from the matrix $\textbf{H}_0(t)$. The practical justification of this flatness is that the bandwidth $B$, is typically significantly smaller than the central frequency $f_\text{c}$ implying small relative variation of $\lambda(\ell)$. Therefore, we will use $\lambda(\ell)=\lambda=v/f_\text{c}$ ($v$ being the speed of light), the wavelength corresponding to $f_\text{c}$, to compute $\textbf{H}_0(t)$ for every frequency subchannel}. The second point is that the directivity gains are assumed to be independent of the antenna indices which is practical as the distance between the plane and the ABS is much larger than the antenna array aperture at the ABS and the plane. As a result, the angle of arrival (AoA) of the signal is almost the same for all antenna elements at the ABS which motivates the independence of the ABS antenna directivity gains from the antenna index. A similar argument can be made for the plane antenna array.

In a similar fashion, we assume that in its generic form, interfering link $i$ is a LoS MISO channel which can be characterized by channel vector $\textbf{h}_i(t) \in \mathbb{C}^{N_\text{P}}$ at time slot $t$. The element $m$ of this vector is defined as (similar to Eq.~\eqref{eq:a2g-chan}, we assume that the variation of $\textbf{h}_i(t)$ over different frequency subchannels is negligible) 
\begin{align}
    [\textbf{h}_i(t)]_m =  \sqrt{\beta_i(f_\text{c},t)G^\text{P}_{i}(t)G^\text{T}_{i}(t)}~e^{-j \frac{2\pi}{\lambda} r_{m,i}(t)}\label{eq:int-chan},
\end{align}
where $\beta_i(f_\text{c},t)$ is the path loss over the interfering link $i$ at time slot $t$ and $r_{m,i}(t)$ is the distance between antenna $m$ of the plane and TBS $i$ at time slot $t$. Additionally, $G^\text{P}_{i}(t)$ is the transmit directivity gain of the plane's antennas, in the direction toward TBS $i$ at time slot $t$. Also, $G^\text{T}_{i}(t)$ denotes the receive directivity gain of TBS $i$ in the direction of the plane at time slot $t$.

\begin{remark} \label{rem:dir-gain}
$G^\text{T}_{i}(t)$ can capture the scenarios where TBS $i$ either has a single (Omni or Dir) antenna or is equipped with multiple antennas. However, it is not straightforward to model the exact directivity gain $G^\text{T}_{i}(t)$ if TBS $i$ is equipped with smart antennas (e.g., massive arrays capable of beamforming or decoding in massive MIMO systems). The reason is that in such cases, this directivity gain depends on the beamforming vector or decoding matrix applied at TBS $i$ which is unknown from the standpoint of the A2G system and varies with time since it is (re-)computed based on the cellular users varying channel vectors. However, as will be seen in Section~\ref{sec:prob-form}, TBS $i$ directivity gain $G^\text{T}_{i}(t)$ is only used in one of the system constraints to limit the A2G interference on TBS $i$. Therefore, it would suffice to find an upper-bound for the interference power --- without knowing the exact receive directivity gain at each time slot --- and then ensure that the upper-bound does not exceed a pre-determined interference power threshold at any time. \textcolor{black}{Fortunately, as the type and number of antenna elements at each TBS in the network are known to the operator, we can use the fact that the receive directivity gain is upper-bounded by the number of antenna elements times the maximum directivity gain per antenna element in all directions in all time slots when a TBS is equipped with an array of identical (Omni or directional) antenna elements~\cite[Chapter 3.3.1]{tse2005fundamentals}}. Therefore, for a TBS equipped with a smart antenna array, we model its receive directivity gain as a uniform gain equal to its number of antenna elements times the maximum directivity gain per antenna element (in all directions) as an upper-bound.  
\end{remark}

We note that Equations \eqref{eq:a2g-chan} and \eqref{eq:int-chan} are generic and one can obtain the corresponding equations for each of the four described scenarios as special cases. For example, in Scenario 1, if the plane and ABS are equipped with a single directional antenna, we have a special case with $N_\text{A}=N_\text{P}=1$ (i.e., the channel matrix will be reduced to a scalar). Furthermore, the directivity gains $G^\text{P}_{i}(t)$ and $G^\text{A}(t)$ depends on the radiation and reception patterns of the plane antenna and ABS antenna, respectively, as well as the relative location of the plane w.r.t. the ABS. 
It is important to note that although our problem formulation and solution are general and works with any given directivity model, we will introduce and use 3GPP directivity model for $G^\text{P}_{0}(t)$, $G^\text{A}(t)$, $G^\text{P}_{i}(t)$, and $G^\text{T}_{i}$ in our numerical evaluations whenever a model is required.

Note that the plane trajectory in its descent path, the locations and characteristics (e.g., type, orientation, radiation and/or reception pattern) of antennas installed on the plane and the ABS, and the maximum directivity gain of each TBS are assumed to be known. This makes it possible to compute $\textbf{H}_0(t)$ and $\textbf{h}_i(t), i\in \mathcal{I}$  for every time slot. In the next section we explain how this information can be used to optimize the transmit power, number of used subchannels, and beamforming at the plane as well as beamforming at the ABS so as to maximize A2G link throughput while guaranteeing a maximum interference power on each TBS at every time slot.

\section{Problem Formulation and Proposed Solution} \label{sec:prob-form}
In this section, we consider the channel described in Section~\ref{sec:sys-model} and formulate our research problem. \textcolor{black}{We develop a solution technique to find an upper-bound and a feasible solution to the problem and show that they are close to each other.} Our problem is generic and, as we will show, can be specialized to each of the four scenarios described in Section~\ref{subsec:scenarios}. 

\subsection{Problem Formulation}
Suppose at time slot $t$, the plane uses $M(t)$ subchannels to transmit signal $s(t)$ to the ABS where $\mathbb{E}[|s(t)|^2]=1$. The plane transmits vector $\textbf{x}(t) = \textbf{w}(t)s(t)$, where $\textbf{w}(t) \in \mathbb{C}^{N_\text{P}}$ is the beamforming (BF) vector applied at the plane antenna array at time slot $t$ and is a control variable. The total expected transmit power of the plane is $\mathbb{E}[||\textbf{x}(t)||^2]=||\textbf{w}(t)||^2$ which is constrained not to exceed the plane's total power budget denoted by $P_\text{max}$. Therefore, there is a sum power constraint (SPC) given by $||\textbf{w}(t)||^2\leq P_\text{max}$. Additionally, since channels are considered to be frequency flat, we assume the transmit power is equally divided among the subchannels used for A2G transmissions at each time slot.  
We also consider per antenna power constraint (PAPC) which restricts the amount of power that can be radiated from each antenna element in the plane's array. \textcolor{black}{This constraint is essential in practical schemes with a power amplifier for each antenna} as it prevents the non-linearity of power amplifiers~\cite{papc2019}. \textcolor{black}{It is straightforward to show that the expected power radiated from antenna element $m$ of the plane's array at time slot $t$ is equal to $|[\textbf{w}(t)]_m|^2$~\cite{papc2018}.} Consequently, we have PAPCs as $|[\textbf{w}(t)]_m|^2\leq P_\text{ant}, m \in\{1,2,\ldots,N_\text{P}\}$, where $P_\text{ant}$ is the maximum allowable radiated power for each antenna element. We assume that the ABS is sufficiently far from the cellular operator's mobile users so that the transmission of the plane over the A2G link is not impacted by the cellular users' uplink transmissions, i.e., there is no interference at the ABS.  Therefore, the ABS receives the vector $\textbf{y}(t)=\textbf{H}_0(t)\textbf{x}(t)+\textbf{n}(t)$, where $\textbf{n}(t)\in \mathbb{C}^{N_\text{A}}$ is the AWGN noise vector with distribution $\mathcal{CN}(\textbf{0}_{N_\text{A}},M(t)b\sigma^2\textbf{I}_{N_\text{A}})$, and $\sigma^2$ is the power spectral density of thermal noise. 

The ABS applies its BF vector $\textbf{v}(t)$, another control variable, and gets $y(t)=\textbf{v}^H(t)\textbf{y}(t)=\textbf{v}^H(t)\textbf{H}_0(t)\textbf{w}(t)s(t) + n(t)$, where $n(t)=\textbf{v}^H(t)\textbf{n}(t)$. The SNR of A2G link at time slot $t$ per channel can be computed using the following lemma.
\begin{lemma}\label{lemma1}
    The SNR of A2G link per subchannel can be written as $\gamma_0(t) = \frac{\left| \widetilde{\textbf{v}}(t)^H \textbf{H}_0(t) \textbf{w}(t) \right|^2 }{   M(t)b\sigma^2 }$, where $\widetilde{\textbf{v}}(t)=\textbf{v}(t)/\|\textbf{v}(t)\|$.
\end{lemma}
The proof of this lemma is provided in Appendix~\ref{a1} where it is shown that the norm of $\textbf{v}(t)$ does not impact the SNR since it affects both signal power and noise power the same way. Therefore, the only unknown parameter at the receiver is the direction of $\textbf{v}(t)$ to be optimized whereas both magnitude and direction of transmitter BF vector $\textbf{w}(t)$ should be optimized. 

We consider a practical scenario where discrete MCSs are used for transmissions.
Let $f(\cdot)$ denote the MCS function of the system. This function is a step-wise non-decreasing function of SNR and has the following generic form $f(\gamma)=e_1\textbf{1}_{[\gamma_1,\gamma_2)}(\gamma) + e_2 \textbf{1}_{[\gamma_2,\gamma_3)}(\gamma)+\cdots + e_{|\mathcal{H}|} \textbf{1}_{[\gamma_{|\mathcal{H}|},\infty)}(\gamma)$
where $\mathcal{H}$ is the set of available MCSs which determines the SNR thresholds $\gamma_i$'s and spectral efficiency levels $e_i$'s measured in bps/Hz. If the per-subchannel SNR of A2G link $\gamma_0(t) $ falls within the interval $[\gamma_i,\gamma_{i+1})$ for some $i\in\{1,2,\ldots,|\mathcal{H}|-1\}$,  then the plane can transmit using the $i$-th highest-order MCS and achieve a spectral efficiency of $e_i$  (bps/Hz). If $\gamma_{|\mathcal{H}|} \leq \gamma_0(t)$, then the plane can transmit with the highest-order MCS resulting in a spectral efficiency of $e_\text{max}\coloneqq e_{|\mathcal{H}|}$. We assume that the block error rate is negligible. Note that different cellular technologies have different MCS functions. An example of the MCS set used in LTE-A systems is given in Table~\ref{18:MCS:3gpp}, where $|\mathcal{H}| = 15$. The resulting data rate of the plane with SNR $\gamma_0(t)$ over $M(t)$ subchannels at time slot $t$ is $r_0(t) = M(t) b f \big( \gamma_0(t) \big)$ (bits/s).

So far, we have developed a model in which we can compute the data rate of A2G link at every time slot, for any given control variables $M(t)$, $\textbf{w}(t)$ and $\textbf{v}(t)$. These variables, however, can be optimized. Given a time interval of duration $T_s$ seconds (or $T_s/\Delta t$ slots) before landing, the objective is to maximize the volume of transmitted data over the A2G link which is equivalent to maximizing the per time slot data rate $r_0(t)$ in each time slot. Since we assume that A2G link re-uses the UL cellular spectrum to restrict the harmful impact of A2G transmissions on possible cellular UL transmissions, we consider a maximum allowable interference power threshold $\delta$ per subchannel for every TBS at each time slot. Note that the location of the plane as well as the channel gains relative to the TBSs change across different time slots and so does the A2G interference at the TBSs.

At time slot $t$, given the inputs $N_\text{sub}$, $b$, $\sigma^2$, $f(\cdot)$, $P_\text{max}$, $P_\text{ant}$, $\delta$, $ \mathbf{H}_0(t)$ --- characterizing the A2G link --- and $\{\mathbf{h}_i(t)\}_{i\in\mathcal{I}} $ --- characterizing the interfering links --- we can obtain the maximum data rate of the plane under per subchannel interference power constrains by solving the following optimization problem: 
\allowdisplaybreaks
\begin{subequations} \label{eq:opt}
    \begin{align} 
        \boldsymbol{\Pi}: &\max_{ \widetilde{\textbf{v}}(t) \in \mathbb{C}^{N_\text{A}},\textbf{w}(t) \in \mathbb{C}^{N_\text{P}}, M(t) }   M(t) bf\big( \gamma_0(t) \big)  \\ 
        &\textrm{s.t.} \ \  \gamma_0(t)  = \frac{ \left| \widetilde{\textbf{v}}^H(t) \textbf{H}_0(t) \textbf{w}(t) \right|^2 }{ M(t)b\sigma^2},\label{c:1} \\
        & \| \textbf{w}(t) \|^2 \leq P_\text{max}, \label{c:2} \\
        & \big| [\textbf{w}(t)]_m \big|^2 \leq P_\text{ant},\ \forall m\in \{1,2,\ldots, N_\text{P} \},  \label{c:3} \\
        & \| \widetilde{\textbf{v}}(t) \| = 1, \label{c:4} \\
        &\frac{\left| \textbf{h}_i^H(t) \textbf{w}(t) \right|^2}{M(t)} \leq \delta, \ i \in \mathcal{I}, \label{int:thresh}\\
        & M(t) \in \mathcal{M},\label{intc}
    \end{align}
\end{subequations}
where $\mathcal{M} \coloneqq \{1, \ldots, N_{\textrm{sub}} \}$. 
Constraint~\eqref{c:1} defines the per-subchannel SNR of the A2G link according to Lemma~\ref{lemma1}, constraints~\eqref{c:2} and~\eqref{c:3}  enforce the SPC and PAPC constraints, constraint~\eqref{c:4} follows the result of Lemma \ref{lemma1}, and finally constraint~\eqref{int:thresh} ensures that per-subchannel A2G interference power at each TBS is below the threshold $\delta$.

Solving the optimization Problem~$\boldsymbol{\Pi}$ allows for the computation of the maximum volume of maintenance data that can be offloaded at every time slot by adaptively modulating the transmitter and receiver beam vectors as well as the number of transmit subchannels at every time slot while the planes moves along the descent path. We note that Problem~$\boldsymbol{\Pi}$  is non-convex even after relaxing the integrality constraint in Eq.~\eqref{intc} because of the non-concavity of the objective function under maximization. This problem can be solved differently depending on the scenario. We first explain the solution technique for the most complex scenario, i.e., Scenario~4 where the plane and ABS are equipped with a UPA, and then show how this solution technique can be simplified for other scenarios.

Our objective is to obtain a `good' feasible solution for Problem~$\boldsymbol{\Pi}$. We will do so by first providing an upper-bound for the optimal objective value of the problem. We then propose a procedure to construct a feasible solution for the problem. Finally, we compare the performance of the constructed feasible solution with the upper-bound to show the quality of that solution.

\subsection{Obtaining an Upper-bound} \label{subsec:upper}
In this section, we propose an approach to find an upper-bound for the optimal objective value of Problem~$\boldsymbol{\Pi}$ by following the steps below.\\

\noindent \textbf{Step 1}:
We first obtain an optimal ABS BF vector $\widetilde{\textbf{v}}(t)$. To this end, we use the fact that the A2G link is  (almost surely) the LoS path. As a result, matrix $\textbf{H}_0(t)$ is (almost surely) rank-one. \textcolor{black}{Furthermore, it is practical to assume that the distance between the plane and ABS is much larger than the antenna array aperture of the plane and ABS.} 
Under this assumption, it can be shown that $\textbf{H}_0(t)$ can be decomposed into
$$\textbf{H}_0(t) = \sqrt{\beta_0(f_\text{c},t) G^\text{P}_{0}(t)G^\text{A}(t)} \textbf{u}_\text{A}(t)\textbf{u}_\text{P}^H(t),$$ with $\|\textbf{u}_\text{A}(t)\|^2=N_\text{A}$ and $\|\textbf{u}_\text{P}(t)\|^2=N_\text{P}$, where $\textbf{u}_\text{A}(t) \in \mathbb{C}^{N_\text{A}}$ and $\textbf{u}_\text{P}(t)\in \mathbb{C}^{N_\text{P}}$ are called the \textit{response vectors} (a.k.a. \textit{steering vectors} or \textit{spatial signatures}) of the ABS and the plane, respectively~\cite{distance,response-vector}. 
Lemma~\ref{lemma2} below uses this model and provides an optimal BF vector for the ABS.
\begin{lemma}\label{lemma2}
    Under the above assumption, there exists an optimal solution to Problem~$\boldsymbol{\Pi}$ in which $\widetilde{\textbf{v}}^{*}(t) = \textbf{u}_\text{A}(t)/\sqrt{N_\text{A}}$. 
\end{lemma}
The proof of this lemma is provided in Appendix~\ref{a2}. Lemma~\ref{lemma2} introduces an optimal choice for BF vector $\widetilde{\textbf{v}}(t)$ which results in a receive BF gain of $N_\text{A}$ at the ABS. The physical interpretation of using $\widetilde{\textbf{v}}^{*}(t)$ is for the ABS to create a narrow reception 
directivity pattern toward the direction of the plane. 
We replace the ABS BF vector $\widetilde{\textbf{v}}(t)$ with its optimal value in Problem~$\boldsymbol{\Pi}$ leading to an equivalent problem with a reduced size which we refer to as $\boldsymbol{\Pi}_1$.\\ 

\begin{figure}[!t] 
\centering
       \includegraphics[scale=0.2]{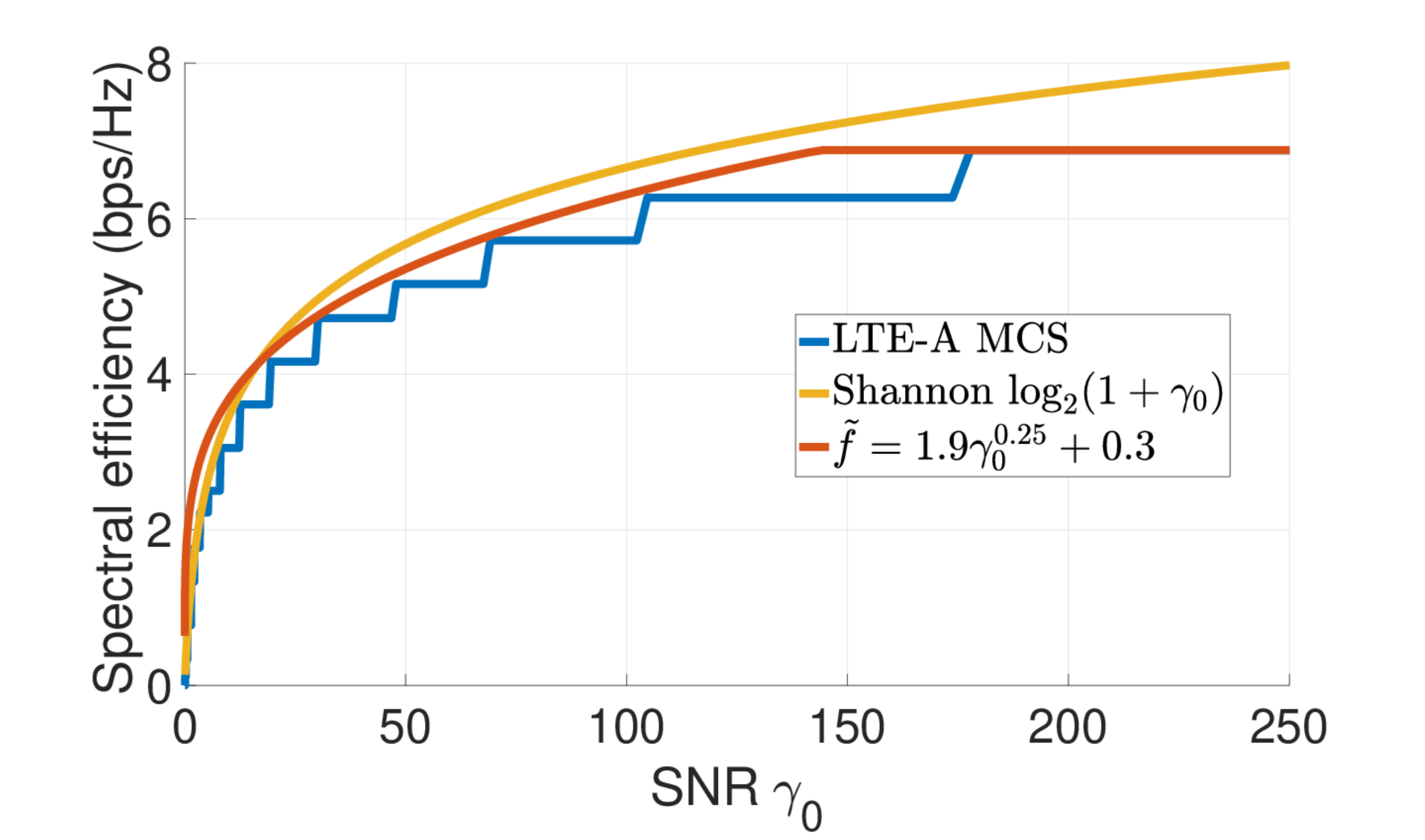}
        \caption{The MCS function: Spectral efficiency achievable with LTE-A as a function of (unitless) SNR.  }
       \label{f:fnc}
\end{figure}

\noindent \textbf{Step 2}:
In the next step, we replace the non-continuous non-concave stepwise MCS function $f(\cdot)$ with a continuous and concave upper-bound represented by $\tilde{f}(\cdot)$. Using this upper-bound instead of $f(.)$ will only increase the objective value, hence helping us to find an upper-bound for the objective value of Problems $\boldsymbol{\Pi}$ (or $\boldsymbol{\Pi}_1$). We consider the generic form $\tilde{f}(\gamma) \coloneqq \min\{ a\gamma^c+d, e_{\text{max}}\}$ and tune the parameters $a$, $c<1$, $d$, and $e_\text{max}$ to fit the MCS function. Note that $e_{\text{max}}$ models the maximum achievable spectral efficiency in the system. For  the set of MCS listed in Table~\ref{18:MCS:3gpp} corresponding to LTE-A, we use the upper-bound $\tilde{f}(\gamma)=\min\{1.9\gamma^{0.25}+0.3,6.88\}$. Fig.~\ref{f:fnc} depicts function $f(\cdot)$ and its concave upper-bound $\tilde{f}(\cdot)$ in this case. Next, we apply the variable transformation $\textbf{W}(t) \coloneqq \textbf{w}(t)\textbf{w}(t)^H $ and reformulate Problem~$\boldsymbol{\Pi}_1$ as follows. 
\begin{subequations} \label{pb:ampl}
    \begin{align} 
       \widetilde{\boldsymbol{\Pi}}_2: & \max_{ M(t), \textbf{W}(t) \in \mathbb{H}^{ N_\text{P} } } \   M(t)  b \Big( a\big(\gamma_0(t)\big)^c + d \Big)  \\ 
        & \textrm{s.t.} \ \gamma_0(t) = \frac{ \textrm{Tr}\big( \textbf{H}^H_0(t) \textbf{U}_\text{A}(t) \textbf{H}_0(t) \textbf{W}(t) \big)}{N_\text{A} M(t) b\sigma^2}, \label{eq:a2gs-snr}\\
        &\gamma_0(t) \leq \left((e_{\text{max}}-d)/a\right)^{1/c}, \label{eq:maxrate}\\
        & \textrm{Tr}(\textbf{W}(t)) \leq P_\text{max}, \\
        & \big[ \textbf{W}(t) \big]_{m,m} \leq P_\text{ant},\ \forall m\in \{1,2,\ldots, N_\text{P} \},  \\
        &  \frac{\textrm{Tr}\big(\textbf{H}_i(t)  \textbf{W}(t) \big)}{M(t)} \leq \delta, \ i \in  \mathcal{I}, \label{eq:interference}\\
        & \textrm{Rank} (\textbf{W}(t)) = 1, \label{eq:rank}\\
        & M(t) \in \mathcal{M},
        \label{integer}
    \end{align}
\end{subequations}
where $\textbf{U}_\text{A}(t) \coloneqq \textbf{u}_\text{A}(t) \textbf{u}_\text{A}^H(t)$, $\textbf{H}_i(t) \coloneqq \textbf{h}_i(t)\textbf{h}_i^H(t),~i\in \mathcal{I}$ and both matrices are known. Note that constraint~\eqref{eq:maxrate} enforces that $\tilde{f}(\gamma_0(t)) \leq e_{max}$. 
Constraint~\eqref{eq:rank} is the direct result of the transformation $\textbf{W}(t) \coloneqq \textbf{w}(t)\textbf{w}(t)^H $ that we used to derive this new optimization problem. 
Furthermore, the optimal objective value of Problem~$\widetilde{\boldsymbol{\Pi}}_2$ is an upper-bound for that of Problem~$\boldsymbol{\Pi}_1$. All the constraints in Problem~$\widetilde{\boldsymbol{\Pi}}_2$ are now convex except the constraints in~\eqref{eq:rank} and~\eqref{integer}. \\

\noindent \textbf{Step 3}: To solve Problem~$\widetilde{\boldsymbol{\Pi}}_2$ in time slot $t$, we first fix $M(t)$ to a value $M\in\{1,\ldots,N_\text{sub}\}$. Next, we relax the problem by removing the rank one constraint \eqref{eq:rank} which would lead to an upper-bound problem. Given a fixed value for $M(t)$, the objective function of Problem~$\widetilde{\boldsymbol{\Pi}}_2$ is concave w.r.t. $\textbf{W}(t)$. Therefore, the new maximization problem, which we refer to as $\widetilde{\boldsymbol{\Pi}}_3(M)$, is a convex problem and can be solved using available platforms such as CVX or AMPL~\cite{cvx}. We solve Problem~$\widetilde{\boldsymbol{\Pi}}_3(M)$ for all possible values of $M\in\{1,\ldots,N_\text{sub}\}$. Let $\widetilde{\mathbf{W}}_3^*(M,t)$ and $\widetilde{O}_3^*(M,t)$ denote the optimal solution and the corresponding optimal objective value in Problem~$\widetilde{\boldsymbol{\Pi}}_3(M)$. It is straightforward to see that $\widetilde{O}_3^*(M,t)$ is an upper-bound for the objective value of Problems~$\boldsymbol{\Pi}$ and $\boldsymbol{\Pi}_1$ for any fixed $M(t)=M \in\{1,\ldots,N_\text{sub}\}$. 
Note that to obtain this upper bound, we have essentially replaced the discrete MCS function by a continuous upper-bound, and relaxed the constraint on the rank of matrix $\textbf{W}(t)$.

In the following, we propose a method to construct a good feasible solution for Problem~$\boldsymbol{\Pi}_1$ (or $\boldsymbol{\Pi}$). The upper-bound obtained in this section will be used to evaluate the performance of that feasible solution.

\subsection{Constructing a Good Feasible Solution for Problem~$\boldsymbol{\Pi}_1$} \label{subsec:feas}
It proved difficult to compute good feasible solutions for Problem~$\boldsymbol{\Pi}_1$ for some of the time slots. This is due to the constraint on the rank of matrix $\textbf{W}(t)$ which we relax for computing the upper-bound but we need to enforce for the feasible solution.

To construct a feasible solution, we follow the same three steps mentioned in Section~\ref{subsec:upper} with minor modifications as follows. In Step 2, instead of using a concave upper-bound $\tilde{f}(\cdot)$ for the actual MCS function $f(.)$, we use another concave approximation which does not have to be an upper-bound and, hence, can be tighter. Note that in Section~\ref{subsec:upper}, we had to use the upper-bound function $\tilde{f}$  as we sought an upper-bound for the objective value of Problem~$\boldsymbol{\Pi}_1$.  However, since we want to find a good feasible solution, we may use any tight approximation for $f$ (including $\tilde{f}$).  
Let $\{\hat{f}_i(.)\}_i$ denote a set of such tight approximations. Similar to Step 2 in Section~\ref{subsec:upper}, we consider the generic form  $\hat{f}_i(.) = \min\{ a_i\gamma^{c_i}+d_i, e_{\text{max}}\}$ with $c_i < 1$. We follow the procedure below to find a feasible solution to Problem~$\boldsymbol{\Pi}_1$ corresponding to each of the approximation functions and then pick the function that gives the feasible solution closest to the upperbound (that we derived above) as the best approximation. 

Pick $\tilde{f}(.)\in \{\hat{f}_i(.)\}_i$. Let Problems~$\widehat{\boldsymbol{\Pi}}_2$ and $\widehat{\boldsymbol{\Pi}}_3(M)$ be the counterparts of Problems~$\widetilde{\boldsymbol{\Pi}}_2$ and $\widetilde{\boldsymbol{\Pi}}_3(M)$ when using approximation $\hat{f}(.)$ instead of $\tilde{f}(.)$. Additionally, let $\widehat{\mathbf{W}}_3^*(M,t)$ and $\widehat{O}_3^*(M,t)$ denote the solution and its corresponding objective value to Problem~$\widehat{\boldsymbol{\Pi}}_3(M)$ which can be found using convex programming. For a given $M$, there are two possibilities when solving Problem~$\widehat{\boldsymbol{\Pi}}_3(M)$:

\subsubsection{$  \text{\textbf{Rank}}(\widehat{\mathbf{W}}_3^*(M,t))=1$} In this case, $\widehat{\mathbf{W}}_3^*(M,t)$ is also an optimal solution to Problem~$\widehat{\boldsymbol{\Pi}}_2$ when $M(t)=M$. Let $\widehat{O}_2^*(M,t)$ be the objective value of of Problem~$\widehat{\boldsymbol{\Pi}}_2$ when $\mathbf{W} = \widehat{\mathbf{W}}_3^*(M,t)$ and $M(t)=M$. Clearly, in this case we have $\widehat{O}_2^*(M,t)=\widehat{O}_3^*(M,t)$. Furthermore, we can expand $\widehat{\mathbf{W}}_3^*(M,t)$ as $(\widehat{\mathbf{w}}^*(M,t))^H\widehat{\mathbf{w}}^*(M,t)$, where $\widehat{\mathbf{w}}^*(M,t)$ is a feasible solution for Problem~$\boldsymbol{\Pi}_1$ when $M(t)=M$. We let $O_1^*(M,t)$ be the objective value of Problem~$\boldsymbol{\Pi}_1$ when $M(t)=M$ and $\mathbf{w}(t) = \widetilde{\mathbf{w}}^*(M,t)$. We will show that this case happens quite often in our numerical examples provided in Section~\ref{results}. 
\subsubsection{$\text{\textbf{Rank}}(\widehat{\mathbf{W}}_3^*(M,t))>1$}
In this case, $\widehat{\mathbf{W}}_3^*(M,t)$ is not a feasible solution for Problem~$\widehat{\boldsymbol{\Pi}}_2$ and finding a global solution for this problem is difficult. However, there are a number of ideas developed in the context of semi-definite relaxation (SDR)~\cite{luo2010semidefinite}, which can be adopted to generate feasible solutions for Problems~$\widehat{\boldsymbol{\Pi}}_2$ and $\boldsymbol{\Pi}_1$ by processing $\widehat{\mathbf{W}}_3^*(M,t)$. We note that since the objective value of Problem~$\widehat{\boldsymbol{\Pi}}_3(M)$ in this case provides an upper-bound for $\widehat{\boldsymbol{\Pi}}_2$, we can use it to evaluate the quality of such feasible solutions. One approach is to take the eigenvector corresponding to the largest eigenvalue of matrix $\widehat{\mathbf{W}}_3^*(M,t)$, and use it as a feasible solution for Problem~$\boldsymbol{\Pi}_1$. Although this simple method is optimal when $\text{Rank}(\widehat{\mathbf{W}}_3^*(M,t))=1$, it is not the best strategy when $\text{Rank}(\widehat{\mathbf{W}}_3^*(M,t))>1$. In this paper, we adopt another approach called the \textit{randomization} technique which can lead to better solutions. This technique has been used along with SDR in several prior works in the MIMO literature such as~\cite{papc2018,papc2019}. The idea is to use $\widehat{\mathbf{W}}_3^*(M,t)$ to generate $N_\text{trial}$ rank-one solutions the best of which (in terms of the objective value of Problem~$\boldsymbol{\Pi}_1$) is then chosen as a feasible solution. The details of our approach is presented in Algorithm \ref{alg:rand}. The output of this algorithm is BF vector $\widehat{\mathbf{w}}^*(M,t)$ which is a feasible solution for Problem~$\boldsymbol{\Pi}_1$. Note that line 10 in Algorithm \ref{alg:rand} ensures the feasibility of the constructed solution. We let $\widehat{O}_1^*(M,t)$ be the objective value of Problem~$\boldsymbol{\Pi}_1$, when $M(t)=M$ and $\mathbf{w}(t) = \widehat{\mathbf{w}}^*(M,t)$. Reference~\cite{papc2018} has studied the impact of $N_\text{trial}$ on the performance of the constructed feasible solution. 
We use the result of that study and choose $N_\text{trial}=100$ in our numerical results in Section~\ref{results}. 
This approach yields good feasible solutions in some cases but can also yield very bad feasible solutions. In such cases, we try a heuristic approach where we use the rank-one matrix obtained for the same value of $M$ for the ``closest'' time slot to $t$ and scale the matrix\footnote{
We scale this matrix, say $\textbf{W}$, as follows. The matrix automatically satisfies the SPC and PAPC constraints but it might not satisfy the  constraints~\eqref{eq:maxrate} and~\eqref{eq:interference}. If it does, then we can use it as is. Otherwise, we replace it with $z\textbf{W}$ where $z>0$ is a scalar chosen so that all the constraints are met.} to obtain a feasible solution to Problem~$\widetilde{\boldsymbol{\Pi}}_2$, and consequently to Problem~$\boldsymbol{\Pi}_1$. We found that this approach yielded good feasible solutions in many cases. Hence, for a given $M$ and a given $t$, we select the solution (out of the two solutions obtained from Algorithm 1 and rank-one matrix from the closest time slot to $t$) that gave the highest objective value to Problem~$\boldsymbol{\Pi}_1$.

\begin{algorithm}
\caption{SDR with Randomization (SDR-R)}
  \begin{algorithmic}[1]
    \STATE Given $M$, solve Problem~$\widehat{\boldsymbol{\Pi}}_3(M)$ and find $\widehat{\mathbf{W}}_3^*(M,t)$.
    \STATE Take the eigenvalue decomposition of $\widehat{\mathbf{W}}_3^*(M,t)$, i.e., $\widehat{\mathbf{W}}_3^*(M,t)=\mathbf{V} \mathbf{\Lambda} \mathbf{V}^H$.
    \IF{$\text{\textbf{Rank}}(\widehat{\mathbf{W}}_3^*(M,t))=1$}
    	\STATE $\widehat{\mathbf{w}}^*(M,t)\gets\mathbf{V}\mathbf{\Lambda}^{1/2}\mathbf{e}$, where $\mathbf{e}=[1,0,0,\ldots,0]^T$.
    \ELSE
    	\STATE $R_{max} \gets 0$.
    	\FOR{$k=1$ to $N_\text{trial}$}
         	\STATE Generate a random vector $\mathbf{e}\in \mathbb{C}^{N_\text{P}}$ whose elements are uniform \textit{i.i.d.} on the unit circle.
     	    \STATE Generate vector $\mathbf{b}=\mathbf{V}\mathbf{\Lambda}^{1/2}\mathbf{e}$ and construct vector
     	     $\mathbf{w}_k$ such that $[\mathbf{w}_k]_i=
     	     [\mathbf{b}]_i/|[\mathbf{b}]_i|
     	     , i\in\{1,2,\ldots,N_\text{P}\}$.
     	    \STATE Generate vector $\mathbf{w}'_k=\sqrt{\ell}\mathbf{w}_k$, where $\ell = \min\{\ell_1,\ell_2,\ell_3\}$,
     	     $\ell_1=\min\{P_\text{ant},P_\text{max}/N_\text{P},\}$, $\ell_2=
     	    \frac{M\delta}{\max_i\{|\textbf{h}_i^H  \mathbf{w}_k|^2\}}$, and $\ell_3= \frac{N_\text{A} M b\sigma^2((e_{\text{max}}+d)/a)^{1/c}}{| \textbf{u}_\text{A}(t) \textbf{H}_0(t) \mathbf{w}_k |^2}$.\\
     	    \STATE Compute A2G SNR $\gamma' = \frac{ | \textbf{u}_\text{A}(t) \textbf{H}_0(t) \mathbf{w}'_k |^2}{N_\text{A} M(t) b\sigma^2}$.
      		\IF {$f(\gamma')>R_{max}$}
         		\STATE $\widehat{\mathbf{w}}^*(M,t) \gets \mathbf{w}'_k$
         		\STATE $R_{max} \gets f(\gamma')$
      		\ENDIF
    	\ENDFOR
    \ENDIF
  \end{algorithmic}
  \label{alg:rand}
\end{algorithm}

At every time slot $t$, we compute $\widehat{O}_1^*(M,t)$, $\forall M\in\mathcal{M}$ 
and choose $M^*(t)=\displaystyle\arg\max_{M\in \mathcal{M}} \widehat{O}_1^*(M,t)$ as the number of  subchannels used at time slot $t$. Furthermore, we pick  $\mathbf{w}(t)  =\widehat{\mathbf{w}}^*(M^*(t),t)$ as the BF vector to be applied at the plane. Consequently, if the plane transmits data during the last $T_s$ seconds before landing, the volume of offloaded data using our feasible solution is 
\begin{align}
   V_{\text{data}} = \sum^{T_s/\Delta t}_{t=1} \widehat{O}_1^*\Big(M^*(t),t\Big)\Delta t, \label{eq:data}
\end{align}
where  the time origin is considered to be $T_s$ seconds before landing. Note that in computing $V_\text{data}$, we assume discrete MCSs are used, hence $V_\text{data}$ is indeed achievable. 
Additionally, to evaluate the performance in Section~\ref{results}, we define the system capacity $V_{\text{cap}}$ as the amount of offloaded data in the same time interval but with $P_{\text{max}}=P_\text{ant}=\infty$ and $\delta=\infty$ which allows the plane to use the best MCS level at every time slot. Hence we have $V_{\text{cap}} = B T_s e_\text{max}$, where $e_\text{max}$ represents the spectral efficiency corresponding to the highest MCS level.  

\subsection{Special Cases}
In the following, we elaborate on how the formulated generic problem and proposed solutions may be simplified to the scenarios described in Section~\ref{subsec:scenarios}.  
\subsubsection{Scenario 1} \label{subsubsec:sce1}
In this scenario, both the plane and ABS have a single Omni or directional antenna and at each time slot, the control variables are the transmit power at the plane as well as the number of subchannels to use. (We assume the tilt and beamwidth of the antenna are given in the case of the directional antenna.) Problem~$\boldsymbol{\Pi}$ can then be reduced to a simplified format which can be solved analytically without going through the described steps in Sections \ref{subsec:upper} and \ref{subsec:feas}. In the following, we show how to find the reduced form of Problem~$\boldsymbol{\Pi}$ in this scenario. Note that the variable $\mathbf{w}(t)$ is a scalar whose squared norm represents the transmit power in this scenario. As a result, we replace $\mathbf{w}(t)$ with $\sqrt{P(t)}$ where $P(t)$ is the plane transmit power at time $t$. Furthermore, we have $\mathbf{\tilde{v}}(t)=1$ since the ABS has a single antenna. Finally, using Eqs.~\eqref{eq:a2g-chan} and \eqref{eq:int-chan}, it can be shown that Problem~$\boldsymbol{\Pi}$ reduces to  
\begin{equation}\label{pb:scenario1}
    \begin{aligned} 
        \boldsymbol{\Pi}_\text{Sce1}:\max_{ M(t)\in \mathcal{M} ,\ P(t) \leq P_\text{max} }  &  M(t) b f\big( \frac{  P(t) \beta_0(f_\text{c},t) G_0^\text{P}(t)G^\text{A}(t)  }{M(t)b\sigma^2} \big)  \\ 
        \textrm{s.t.} \ \ 
        & \frac{P(t)G^\text{P}_i(t)G^\text{T}_i(t)}{M(t)}  \leq \delta, \ i \in \mathcal{I},
    \end{aligned}
\end{equation} 
noting that the optimization variables are the number of subchannels to use and transmit power of the plane. 
The following lemma exploits the structure of this problem and the fact that MCS function $f(\cdot)$ is non-decreasing to provide the closed form solution to this problem. 
\begin{lemma} \label{lem:sce1}
Let tuple $(M^*(t),P^*(t))$ denote the solution to Problem~$\boldsymbol{\Pi}_\text{Sce1}$. We define
$Q^*(M,t) \coloneqq \min\{Q_1(M,t),P_\text{max}\}$, where $Q_1(M,t)=\frac{M\delta}{ \max_{i\in \mathcal{I}}\{G^\text{P}_i(t) G^\text{T}_i(t)\}}$ for $M\in \mathcal{M}$,~and~~ $R(M,t)\coloneqq Mbf\left(\frac{ Q^*(M,t)\beta_0(f_\text{c},t) G_0^\text{P}(t)G^\text{A}(t)}{Mb\sigma^2}\right)$ for $M\in \mathcal{M}$. We have $M^*(t) = \arg\max_{M\in \mathcal{M}}\{R(M,t)\}$ and $P^*(t) = Q^*(M^*(t),t)$. 
\end{lemma}
The proof of this lemma is provided in Appendix~\ref{app:lem-sce1}.

According to Lemma \ref{lem:sce1}, at each time slot $t$, it is sufficient to compute $R(M,t)$ for $M\in \mathcal{M}$ and choose the $M$ leading to the highest value of $R(M,t)$ as the optimal number of subchannels $M^*(t)$. Furthermore, the optimal transmit power $P^*(t)$ is obtained by evaluating $Q^*(M,t)$ for $M=M^*(t)$. We note that the time complexity of this procedure is linear in the number of subchannels $N_\text{sub}$ and that we can solve Problem~$\boldsymbol{\Pi}_\text{Sce1}$ to optimality without having to replace the discrete MCS function in the objective with a concave function.

\subsubsection{Scenario 2}  \label{subsubsec:sce2}
In this scenario, the plane has a single Omni or directional antenna whereas the ABS has an antenna array. As a result, the optimization variables are the number of subchannels to use, the transmit power of the plane, and the BF vector to be applied at the ABS array. We can find an optimal BF vector for the ABS by using Lemma \ref{lemma1}. 
Note that in this scenario $\mathbf{H}_0(t)$ reduces to a column vector of size $N_\text{A} \times 1$ and \textcolor{black}{$\mathbf{u}_\text{A} = \mathbf{H}_0(t)/\sqrt{\beta_0(f_\text{c},t)G^\text{P}_0(t)G^\text{A}(t) }$} in Lemma \ref{lemma2}. It is straightforward to show that after substituting  $\widetilde{\textbf{v}}^{*}_0(t)$ provided by Lemma \ref{lemma2}  in this scenario, Problem~$\boldsymbol{\Pi}$ is equivalent to that for Scenario 1 where the directivity gain of the ABS $G^\text{A}(t)$ can be computed based on the location of the ABS and plane at each time slot. If the UPA elements at the ABS are Omni, $G^\text{A}(t)$ can be further reduced to $N_\text{A}$ (cf. Remark~\ref{rem:dir-gain}) which is the result of proper BF at each time slot. Therefore, Problem~$\boldsymbol{\Pi}$ for Scenario 2 will reduce to Problem~$\boldsymbol{\Pi}_\text{Sce1}$. Subsequently, Lemma \ref{lem:sce1} provides the optimal number of subchannels and  transmit power in Scenario 2 which then can be solved to optimality.

\subsubsection{Scenario 3} \label{subsubsec:sce3}
In this scenario, the ABS has a single Omni or directional antenna whereas the plane is equipped with an antenna array.
The optimization variables in this case are the BF vector to be applied at the plane and the number of subchannels to use. The approach proposed in Section~\ref{subsec:upper} can be used to find an upper-bound and the Section~\ref{subsec:feas} can be used to construct a feasible solution. However,
since the ABS has a single antenna, Step 1 of the proposed approach in Section~\ref{subsec:upper} is not required. Steps 2 and 3 in that section can be applied with minor modifications. Note that in this scenario, $\mathbf{H}_0(t)$ reduces to a row vector of size $1\times N_\text{P}$. Also, as the ABS has a single antenna, it is straightforward to show that in this case we have $\mathbf{U}_\text{A}(t) = \mathbf{I}_{N_\text{A}}$ in Problems~$\widetilde{\boldsymbol{\Pi}}_2$ and $\widehat{\boldsymbol{\Pi}}_2$. 

\subsubsection{Scenario 4} \label{subsubsec:sce4}
In this scenario, both the ABS and  plane are equipped with an antenna array. The optimization variables are the BF vectors to be applied at the plane and ABS and the number of subchannels to use. Consequently, every step of the proposed approach in Section~\ref{subsec:upper} is required to find a performance upper-bound. Also, the method proposed in Section~\ref{subsec:feas} can be used to construct a feasible solution. 


\begin{table*}[!t]
     \caption{MCS set proposed for LTE-A in~\cite{18:MCS:3gpp} 
     }
    \hspace*{-5.5em}
    \label{18:MCS:3gpp}
    \begin{tabular}{|c|c|c|c|c|c|c|c|c|c|c|c|c|c|c|c| }
      \hline 
      \specialcell{ SNR threshold $\gamma$ in dB }& -9.8 & -6.1  &  -2.2 &  1.6 & 3.4 & 5.4 & 7.2 & 9.1 & 11.0 & 12.9 & 14.8 & 16.8 & 18.4 & 20.2 & 22.5  \\ \hline
        \specialcell{ Spectral efficiency $f(\gamma)$ in bps/Hz}&  0.11   &  0.33   &  0.77    & 1.33    & 1.77    & 2.22 &   2.50   & 3.05   &  3.61   &  4.16  &  4.72  & 5.16   &  5.72    &  6.27   &  6.88   \\ \hline 
    \end{tabular}
\end{table*}



\section{Numerical Results}\label{results} 


We consider an airport with a runway of length 4~Km, which is a standard runway length in many international airports (e.g., Heathrow~\cite{heath:runway}). The ABS is placed in the middle of the runaway.  We consider a plane approaching the airport on a descent path with a pitch angle $\psi=\ang{3}$. The plane descends using the Continuous Descent Approach, which is a descent approach used at major airports such as Heathrow~\cite{heath:runway}, at a constant vertical velocity of $v_y = -12.7 $ m/s. The plane starts transmitting the collected maintenance data $T_s$ seconds before landing. The top view of the system is depicted in Fig.~\ref{diag} which shows the location of the ABS as well as the location of 120 TBSs generated randomly near the airport and descent path.  In major airports, such as Heathrow, planes arrive and descend sequentially at the descent path typically with an inter-arrival time of 2 to 5 minutes during busy hours. The descent time for each plane is typically at least 15 minutes. In an online A2G system where the operator wants to maximize the offloaded data volume from the descending planes over a day, a greedy scheduler at the ABS would schedule the planes sequentially when they near the ABS (i.e., when they see the lowest path loss), allowing each plane to transmit for a contiguous period of 2 to 5~minutes (depending on the separation time between the scheduled plane and the preceding one) until the plane lands and leaves the descent path. Following this logic, we consider a range of values for $T_s$ in the interval $[0,300]$ seconds since 300 seconds (5~minutes) would be a realistic value for the maximum allowable transmission time per plane in an online system during busy hours. 

The cellular network operates in FDD mode. To evaluate the potential of both microwave and mmWave frequency bands, we consider two frequency bands, one centered around $f_c=2$ GHz with a bandwidth of $B =20$ MHz and one around $f_c=28$ GHz with a bandwidth of $B=1$ GHz. 
We adopt the LTE-A standard and set the subchannel width and time slot duration to 180 KHz and 1 ms, respectively. 
We set the default per subchannel interference power threshold  $\delta$ to -100 dBm, although we also consider a conservative value, namely -120 dBm, to see the impact of $\delta$ on system performance.  
  
We use the 4G LTE-A MCS set (in Table~\ref{18:MCS:3gpp}), where the highest-order MCS is 256-QAM, as the default one in all our simulations. 
In computing an upper-bound for Problem~$\boldsymbol{\Pi}$ (or equivalently~$\boldsymbol{\Pi}_1$) as described in Section~\ref{subsec:upper}, we consider the upper-bound function $\tilde{f}(\gamma)=\min\{1.9\gamma^{0.25}+0.3,e_\text{max}\}$ with $e_\text{max}=6.88$. 
For deriving a feasible solution as described in Section~\ref{subsec:feas}, we consider a set of four approximation functions $\{\hat{f}_i(\cdot)\}_{i=1}^{4}$ (namely, $ \hat{f}_1(\gamma) = \min\{1.9\gamma^{0.25}+0.3,e_\text{max}\},~\hat{f}_2(\gamma) = \min\{1.9\gamma^{0.25}+0.3,e_\text{max}\},~\hat{f}_3(\gamma) = \min\{ 4.0476 \gamma^{0.185} - 2.4405,e_\text{max}\},~\hat{f}_4(\gamma) = \min\{0.93 \gamma^{0.4} + 0.3,e_\text{max}\}, \min\{ \gamma^{0.37} + 0.3,e_\text{max}\} $). We found these functions by trial and error. One of the functions in the set is the upper-bound function $\tilde{f}$ and the rest of them fall below this function.

The reason why we use the LTE-A MCS function is because we could not find a table with higher-order MCSs for 5G in the literature. 
To be able to understand whether adding higher-order MCSs to the LTE-A table could bring any gains at all, we decided to also use Shannon's formula --- which allows us to quantify the maximum volume of offloaded data achievable in theory --- in our simulations, i.e., the concave and unbounded function $f(\gamma)=\log_2(1+\gamma)$. 

To model the directivity gain of directional antennas at the BSs, we adopt the 3GPP standard's 3D tri-sector model in~\cite{sect:3gpp}. 
The directional gain consists of two components: the azimuth and elevation gains. The azimuth gain is given by 
$ A_{0,H}(t) = -\min \{ 12(\frac{ \epsilon_{0,H}(t,\vec{b}_0)  }{ \ang{65} })^2, 20 \} $ dBi, where \ang{65} is the 3-dB beamwidth on the horizontal plane and $\epsilon_{0,H}(t,\vec{b}_0)$ is the azimuth angle between the plane and the boresight vector $\vec{b}_0$\footnote{Boresight vector of an antenna splits the 3-dB beamwidth equally in two and characterizes the orientation of the antenna.} of the ABS and, as the plane moves, it varies with time. Similarly, the elevation gain is given by 
$A_{0,V}(t) = -\min \{ 12(\frac{ \epsilon_{0,V}(t,\vec{b}_0) - \eta_0  }{ \ang{7} })^2, 20 \}$ dBi, 
where $\epsilon_{0,V}(t,\vec{b}_0)$ is the elevation angle between the plane and the boresight vector $\vec{b}_0$ of the ABS, $\eta_0$ is the tilt angle and \ang{7} is the 3-dB beamwidth on the vertical plane. 
The reception gain is given by
$G^\text{A}(t) = 10^{\big(\Phi_0 -\min \{- A_{0,V}(t) - A_{0,H}(t) , 20 \}\big) / 10 },$
where $\Phi_0$ represents the boresight gain\footnote{Boresight gain is the antenna gain along the boresight vector and is the highest gain the antenna can induce.}.  
The transmitter gain of the plane $G^P_0(t)$ can also be modelled similarly when the plane is equipped with a directional antenna. For the ABS, we assume the BS has a single sector which boresight vector is parallel to and facing the descent path. Furthermore, in all our simulations, whenever we use directional antennas at a TBS, we ensure that the boresight of one of the three sectors is oriented such that it is perpendicular to the projection of the descent path on the ground (x-axis). We do this to ensure that at least one sector at each TBS is always facing the descent path and, hence, all TBSs are always prone to A2G interference.



We assume any (type of) antenna installed on the plane is of pico scale (both in size and antenna gain) since we anticipate that the operator might prefer to use a smaller and cheaper antenna for the plane. For the BSs, we assume the antennas are of macro scale. 
We assume the directional antennas used on the plane and BSs to have a boresight gain of 8 and 17.7 dBi, respectively. 
We consider two values for the plane's transmit power budget $P_\text{max}$ of 1 W which is the standard power budget used for pico antennas in LTE-A, and 40~W which is a hypothetical one. Moreover, for all UPA transmitters, we set the per antenna power constraint $P_\text{ant}$ to 0.2~W.   
We use a 5-by-5 UPA at the plane and a 32-by-32 UPA at the ABS (i.e., $N_\text{P}=25$ and $N_\text{A}=1024$). We keep the size of the UPA at the plane small ($N_\text{P}=25$) since it might not be feasible to install a large UPA on the plane.  As recommended by 3GPP~\cite{3gpp:upa:element,3gpp:2}, we assume each antenna element has a 3D directional radiation pattern with a maximum gain of 8~dBi.

We use the deterministic free-space path loss model in~\cite{3gpp:free:space} along with an added absorption and damping factor $L\big( d_i(t),f_c \big)$ (in dB/Km) given by
$ 32.5 + 20\log_{10}\big( \max\{d_i(t),75\} \times (f_c / 1000) \big) + L\big(d_i(t),f_c \big) \times d_i(t)/1000,  \ \text{[dB]}, $ where $d_i(t)$ is the distance between the plane and BS $i$ at $t$ in meter, and carrier frequency $f_c$ is in MHz. We set $L$ to 0.01 dB/Km for the 2-GHz carrier and to 0.1 dB/Km for the 28-GHz carrier~\cite{absorption}. The rest of the simulation parameters are summarized in Table~\ref{params}. 

There are many possible cases to consider depending on the frequency band being used and the type of cellular system which coexist with the A2G service. We consider two practical cases in the subsequent sections and show in each case how much data can be offloaded for Scenarios 1-4. The cases are:
\begin{itemize}
    \item Case~1: Microwave Band ($f_c=2$ GHz) with directional TBSs.
    \item Case~2: Millimeter Band ($f_c=28$ GHz) with Outdoor UPA-Based  TBSs. 
\end{itemize}
In Case~2, we assume all the BSs (ABS and TBSs) are UPA-based since we are not sure if directional or omni antennas can be deployed in the mmWave band yet. Therefore, we only study Scenario 4 for that case. Furthermore, in studying the two cases --- characterized by a carrier frequency, associated bandwidth, and type of transmitter/receiver antenna  --- we are particularly interested in knowing which one of the following factors limits the performance: A2G interference threshold ($\delta$), MCSs offered by LTE-A ($f(\cdot)$), or transmit power budget ($P_\text{max}$).



\subsection{Quality of the Feasible Solution}
Before presenting the results for each case, we first show the quality of the feasible solution to Problem~$\boldsymbol{\Pi}$ that we derived in Section~\ref{subsec:feas} for each time slot. First recall that there is only a need to do so for Scenarios~3 and 4 since for the first two scenarios, the solution that we obtain is optimal.

\begin{figure}
    \centering
    \includegraphics[scale=0.2]{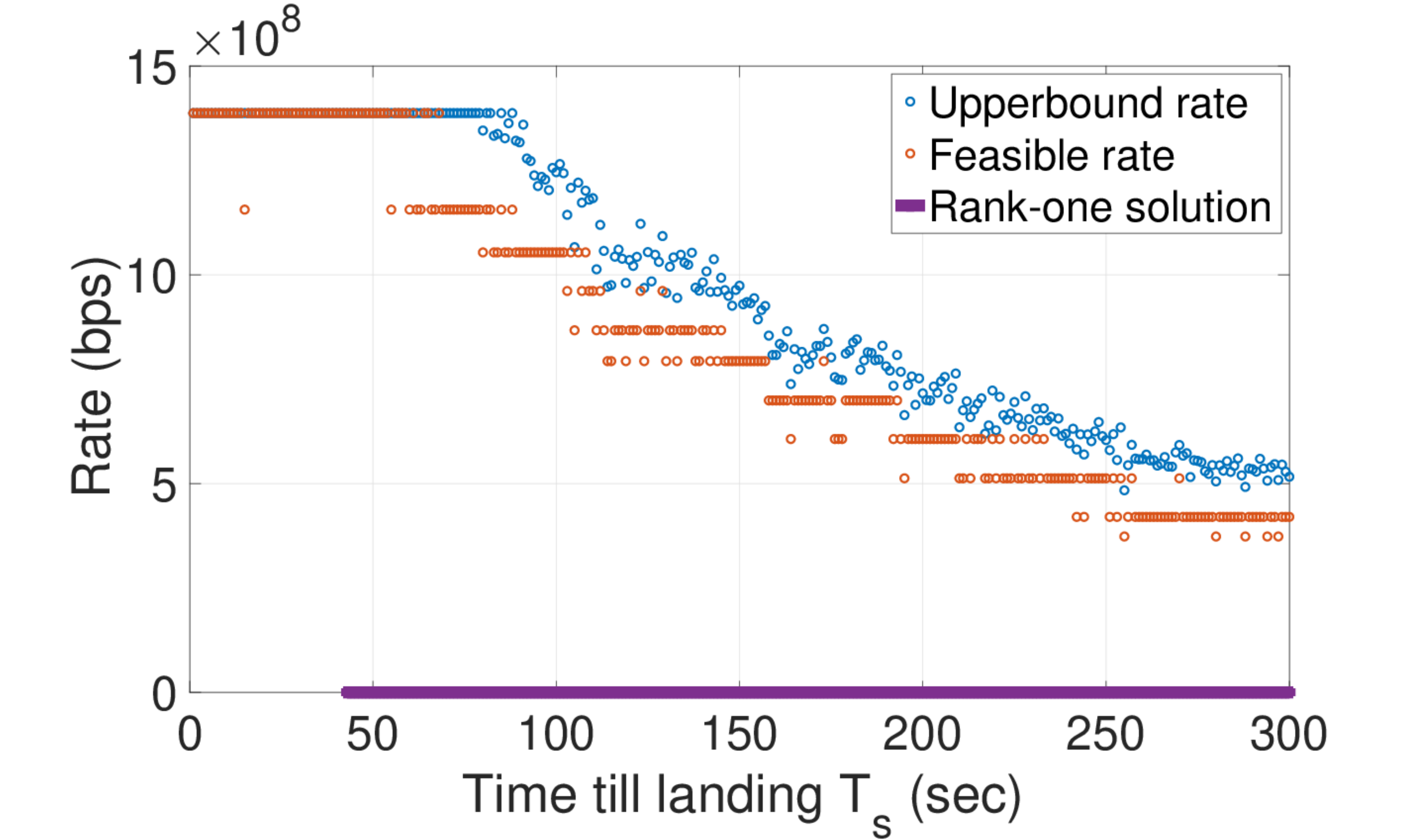}
    \caption{Comparison between the upper-bound and feasible data rate functions corresponding to Problem~$\boldsymbol{\Pi}$ for Case~2 ($P_\text{max}=40$ W, every TBS is equipped with a 16-by-16 UPA).}
    \label{rank:relax}
\end{figure}

To evaluate the quality of the feasible solution for Scenarios~3 and 4 for a given time slot, we first find the upper-bound data rate (bps) for Problem~$\boldsymbol{\Pi}$ as described in Section~\ref{subsec:upper}. We then derive a feasible data rate (using the approximation functions mentioned before). Note that the optimal objective value for Problem~$\boldsymbol{\Pi}$ falls in between these two rates. 
In our simulations, we noticed that this gap is very small for Case~1. 
This is because in the microwave band we can almost always derive a feasible solution that results in a spectral efficiency equal or very close to $e_\text{max}$ (corresponding to the highest-order MCS) for LTE-A MCS function for Scenarios 3 and~4. 
Therefore, we only show the gap for Case~2
when every TBS is equipped with a 16-by-16 UPA.
Specifically, in Fig.~\ref{rank:relax} we show the upper-bound  and feasible data rate functions (in bps) as a function of time till landing $T_s$.  
We also show, with colour purple on the x-axis, the time slots in which the feasible solution matrix $\textbf{W}(t)$ was of rank~one and hence, we did not have to use the algorithm described in Section~\ref{subsec:feas}. 
The figure shows that the rates are  the lowest at the time when the plane is farthest away from the ABS, i.e., $T_s=300$ seconds. As the plane nears the ABS for landing which occurs at $T_s=0$ seconds, the rates gradually increase.
The gap between the upper-bound and feasible solution is less than 10\% in the first  258 seconds and almost zero for the remainder of the descent time. The fact that the solutions in the first 258 seconds are all rank-one means that the 
gap in that interval is due to the gap between the continuous upper-bound function $\tilde{f}$ and the step-wise MCS function $f$. 
In the last 42 seconds, the upper-bound and feasible rate functions overlap since the feasible solution that we derive is good enough to yield the highest spectral efficiency  $e_\text{max}$ (corresponding to the highest-order MCS) which is also why the curves are flat in that interval. 
We should  stress that in finding a feasible solution in the last 42 seconds, the SDR algorithm gave very bad solutions at times and so we had to resort to using the feasible solutions in the previous or subsequent time slots to improve the feasible solutions as described in Section~\ref{subsec:feas}.

\subsection{Case~1: Microwave Band with directional TBSs}
In this case, the system operates in the microwave band. Every TBS is equipped with three directional antennas which is the typical case in LTE-A macro cellular systems. We assume that TBSs antennas have a $\ang{0}$ tilt. Furthermore, the plane and ABS may be equipped with a directional or omnidirectional antenna or a UPA. Therefore, the problem reduces to one of the four scenarios described in Section~\ref{subsec:scenarios}. In the following, we present the results for these scenarios.

\subsubsection{Scenario 1} The plane and ABS are equipped with directional antennas. (Due to lack of space, we do not present our results for Omni antennas since the conclusions remain the same). Additionally, while the plane antenna has no tilt, it is assumed that the ABS is tilted upward by \ang{3} (equal to the pitch angle $\psi$) from the ground so as to have the plane in its line of sight and aligned with its boresight vector. We use Lemma \ref{lem:sce1} to find the optimal transmit power and the number of subchannels at each time slot. Fig.~\ref{2ghz:sec:sec}  shows the offloaded data (in gigabyte) achievable as a function of $T_s$ (time till landing) for two values of the transmit power budget $P_\text{max}$ for the LTE-A MCS and Shannon's function. It also shows the system capacity $V_\text{cap}$, i.e., the maximum achievable volume of data that could be obtained with infinite power budget and no SPC, PAPC, or interference constraints for that MCS function.  
The y-coordinate of each curve is the total data that can be offloaded from the plane accumulated over the transmission period which starts at the corresponding x-coordinate and ends upon landing (i.e., the last $T_s$ seconds of descent). 
The figure shows that the plane can offload around 2~GB of data in the last 300 seconds of descent for both values of the power budget. Hence, a large power budget does not help with improving the performance in that case. This is due to both the weak directionality of the directional antenna at the plane, which makes the system highly susceptible to A2G interference and prevents the plane from using all of its power budget, and the low reception gain of the directional antenna at the ABS. We also note that the results are very far from $V_\text{cap}$, indicating that the system is interference-limited.  
An important observation is that the offloaded data $V_\text{data}$ increases almost linearly with $T_s$, indicating that a significant gain in offloaded data can be achieved by allowing a longer transmission time for the plane. 
The reason for this is that since the system is operating in the microwave band and the plane is in the line of sight of the ABS, the plane transmitting far away from the ABS can still achieve a good enough (non-zero) spectral efficiency thanks to the good channel conditions. 
This observation has an important implication for online A2G systems where a sequence of planes may be on the descent path at the same time. For such a system, scheduling multiple planes simultaneously may bring significant gains (if the inter-plane interference is managed properly). As we will see later, this is not necessarily the case in the mmWave band. The figure also shows the results for when an ideal MCS function (represented by Shannon's formula) is used. Clearly, the system is not MCS-limited, i.e., it would not benefit greatly from higher-order MCSs. 

\subsubsection{Scenario 2}
The plane is equipped with a directional antenna and the ABS is equipped with a 32-by-32 UPA. We use the approach described in Section~\ref{subsubsec:sce2} to find the optimal receiver BF vector at the ABS as well as the transmit power of the plane and the number of subchannels to use. As Fig.~\ref{2ghz:upa:sec} shows, this scenario performs better than Scenario 1 because of the UPA installed at the receiver and improved reception gain. The operator can offer an airliner just above 4~GB of offloaded data in the last 300 seconds of descent for both values of the power budget. Hence, a large power budget is not necessary in this case either. The gap w.r.t.  $V_\text{cap}$ is much smaller than that in Case~1 thanks to the higher reception gain of UPA at the ABS (than a simple directional antenna), indicating that the A2G system is not as much limited by the interference constraint. 
As before, the offloaded data increases linearly with $T_s$. The figure also shows the results for when an ideal MCS function (represented by Shannon's formula) is used. 
The system is MCS-limited, particularly for low values of $T_s$. Hence, it would benefit from higher-order MCSs. 


\subsubsection{Scenario 3}
The plane is equipped with a 5-by-5 UPA and the ABS is equipped with a directional antenna tilted upward by $\ang{3}$ w.r.t. the ground. We use the approach described in Section~\ref{subsubsec:sce3} to find a feasible BF vector at the plane and the number of subchannels to use. Fig.~\ref{2ghz:sec:upa} shows that there is some gain w.r.t. Scenario~2, particularly for smaller values of $T_s$. The offloaded data is now very close to $V_\text{cap}$ which indicates that the system is practically not interference-limited. It is also not power-limited since the results for 1~W are about the same as for 40~W. 
The figure also shows results for Shannon's formula. Clearly, the system is MCS-limited and would greatly benefit from higher-order MCSs, especially when the power budget is 40~W.
\subsubsection{Scenario 4} 
\begin{figure}[!t]
         \centering
         \subfigure[Scenario 1]{
         \includegraphics[scale=0.15]{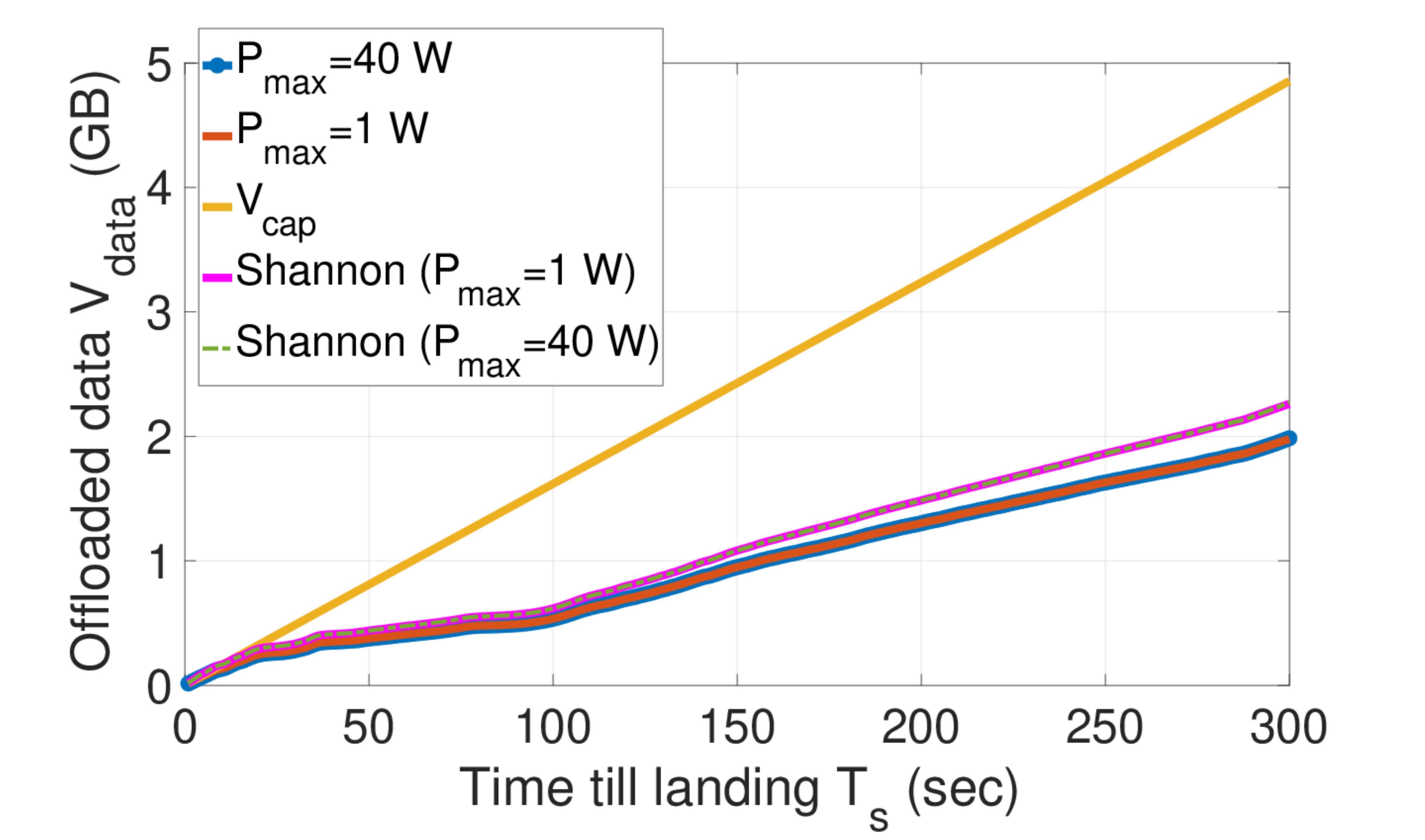} \label{2ghz:sec:sec}}
         \subfigure[Scenario 2]{
         \includegraphics[scale=0.15]{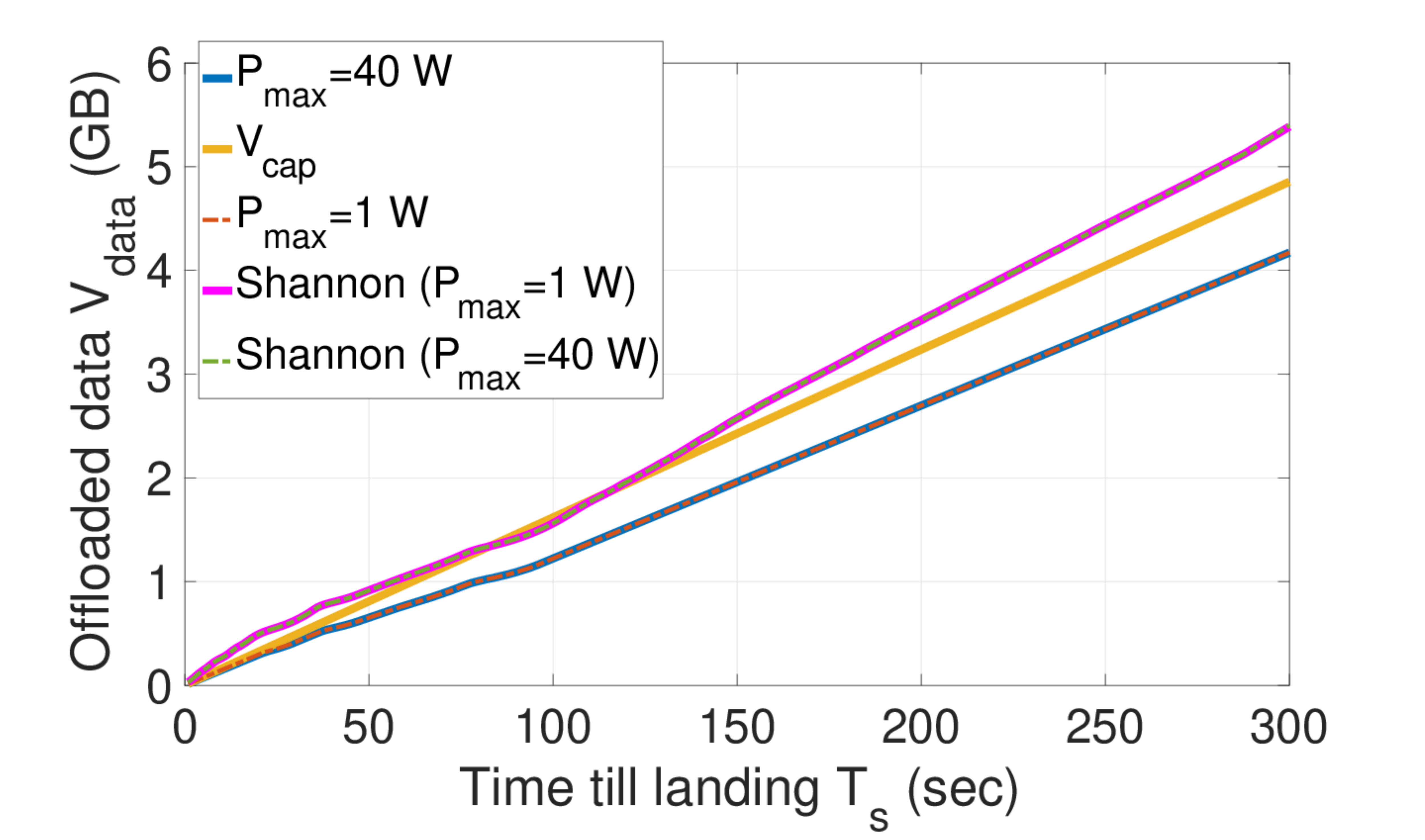} \label{2ghz:upa:sec}}
         \subfigure[Scenario 3]{
         \includegraphics[scale=0.15]{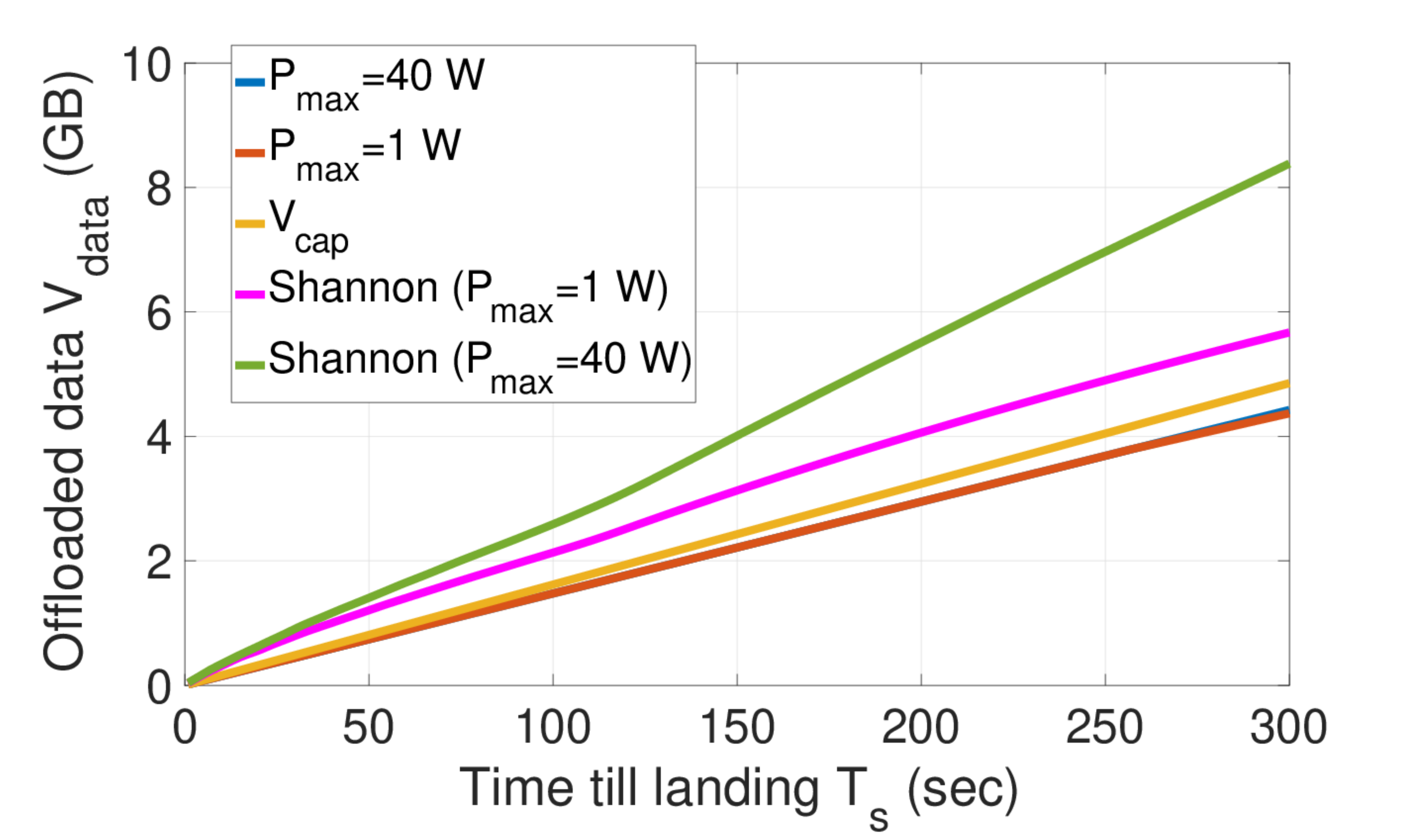} \label{2ghz:sec:upa}}
         \subfigure[Scenario 4]{
         \includegraphics[scale=0.15]{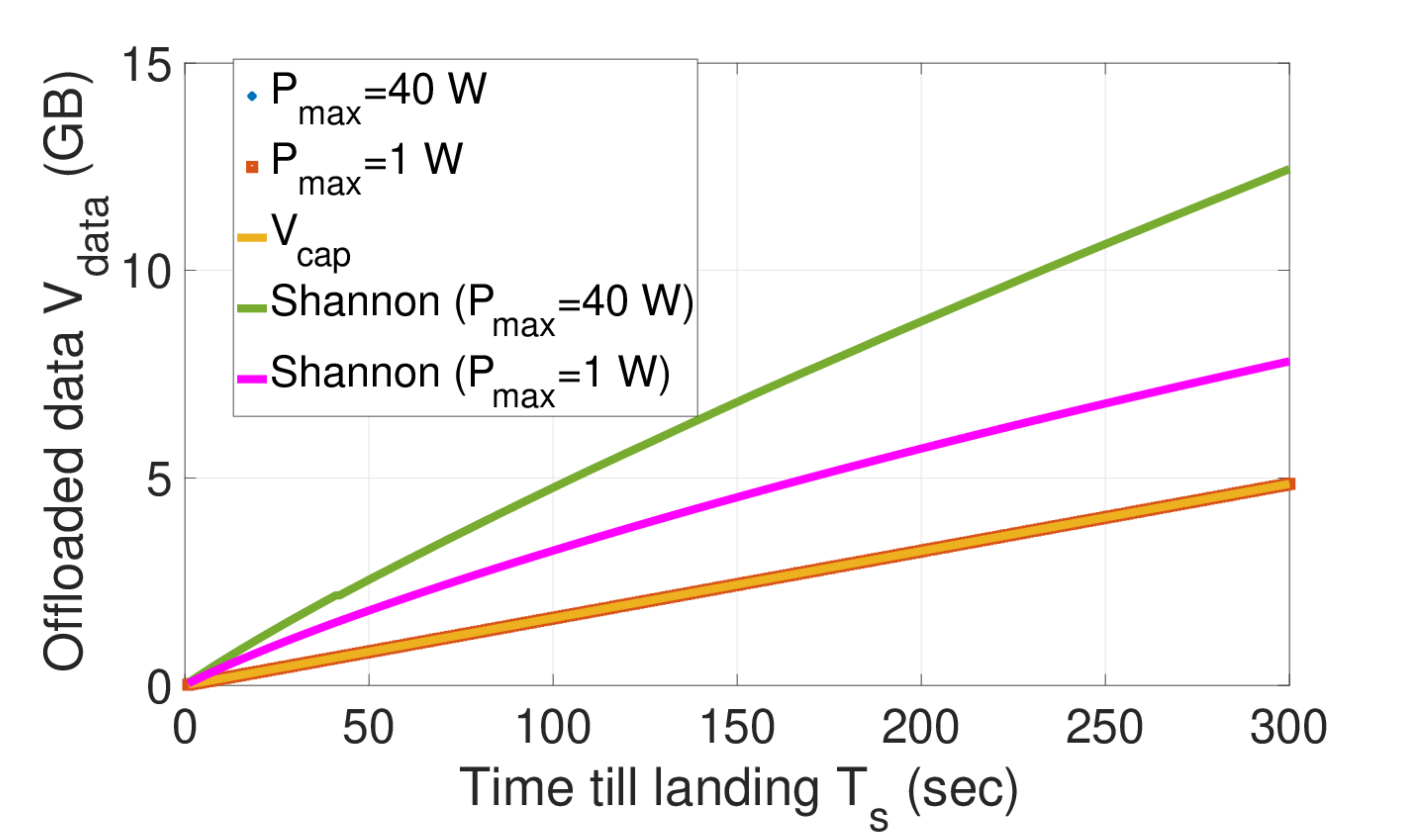} \label{2ghz:upa:upa}}
         \caption{Case~1 --- Scenarios 1-4: Impact of power budget on performance.
         }  \label{figtoto}
    \end{figure}

The plane and ABS are both equipped with a UPA. Fig.~\ref{2ghz:upa:upa} shows results similar to the previous case. 
The performance gain w.r.t. Scenario~3 can be as high as 10\%. 
The figure also shows results for Shannon's formula. Clearly, the system is MCS-limited and would greatly benefit from higher-order MCSs for both power budgets.
  
\begin{figure}[!t]
     \centering
     \subfigure[Scenario 1]{
     \includegraphics[scale=0.15]{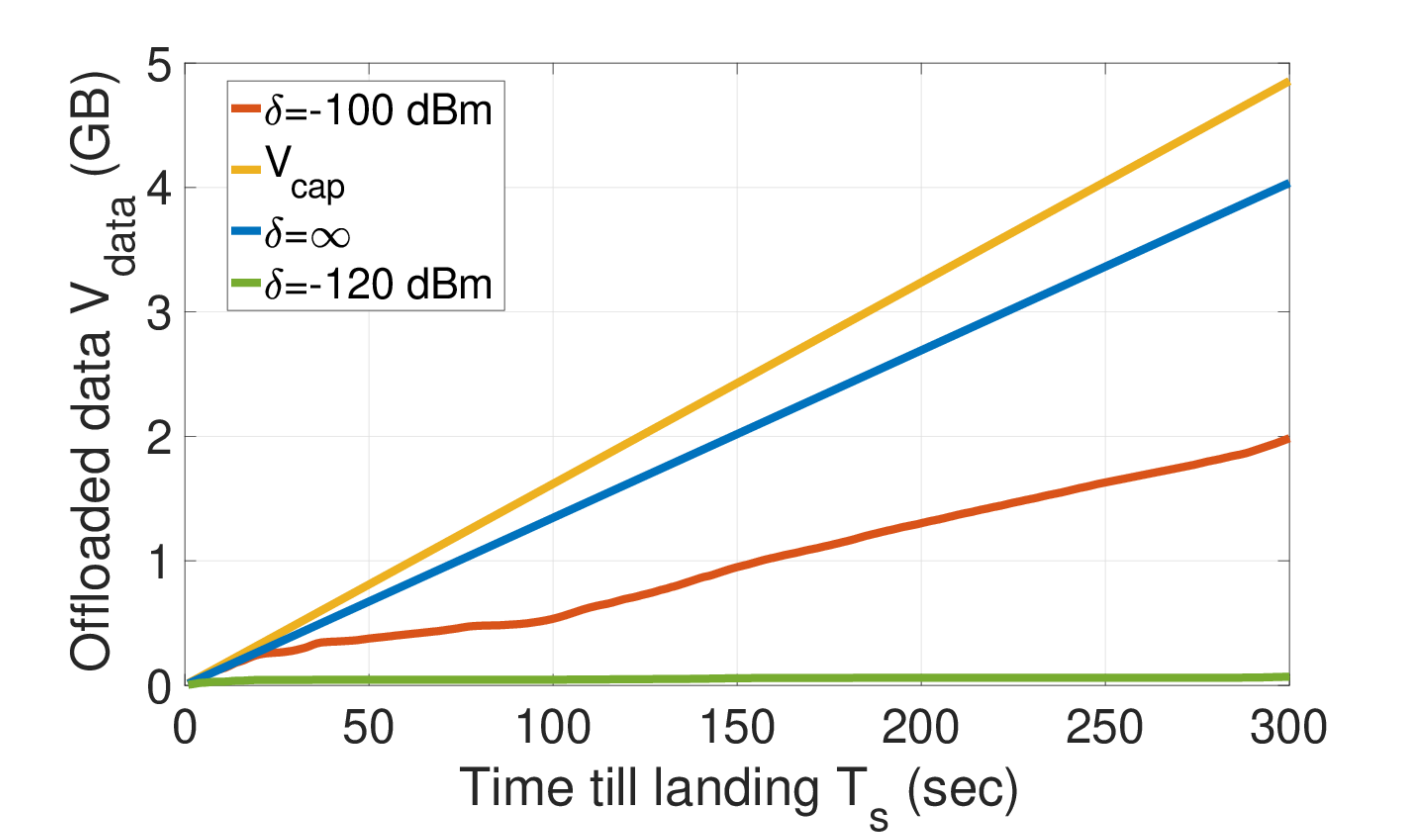} }
     \subfigure[Scenario 2]{
     \includegraphics[scale=0.15]{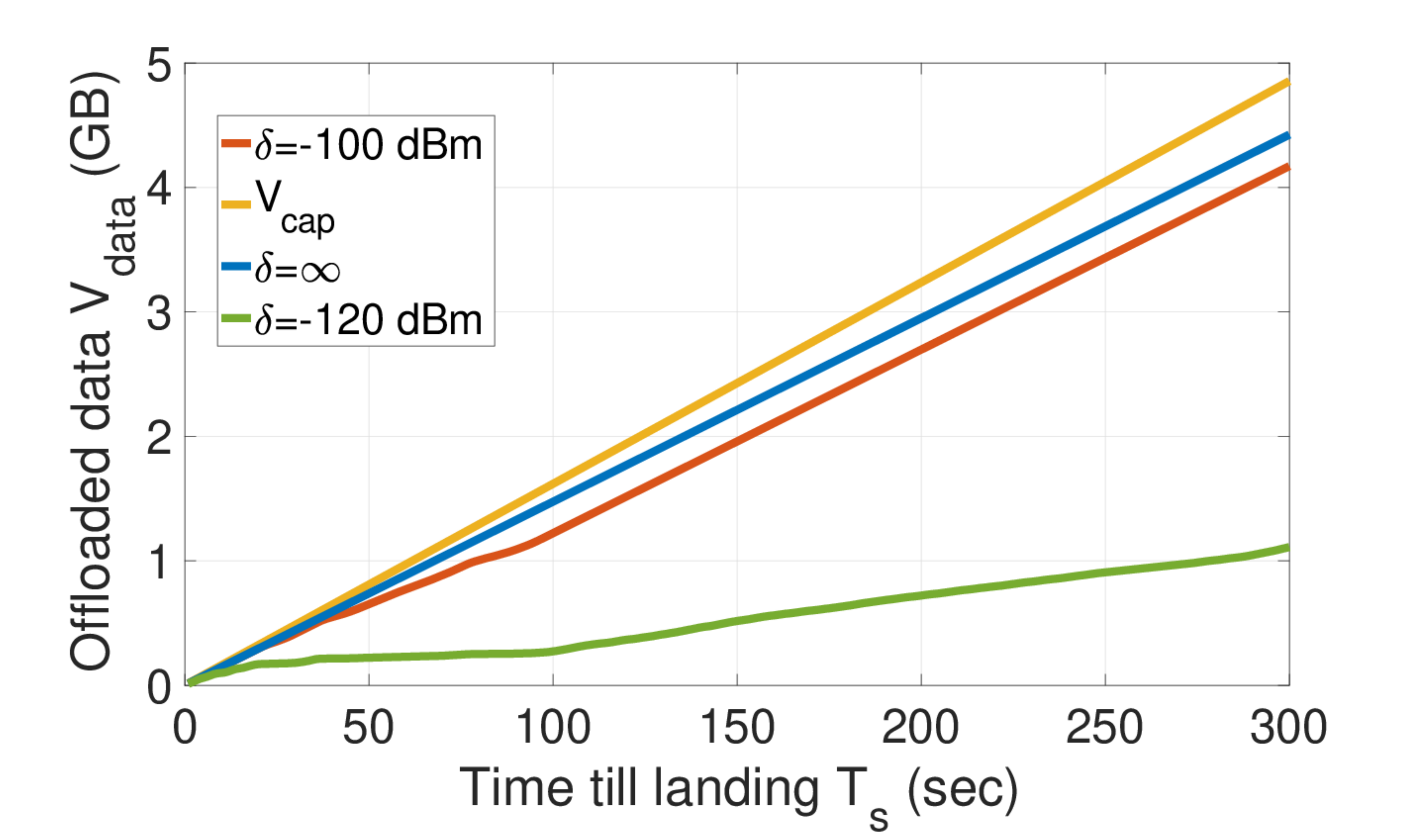} }
     \subfigure[Scenario 3]{
     \includegraphics[scale=0.15]{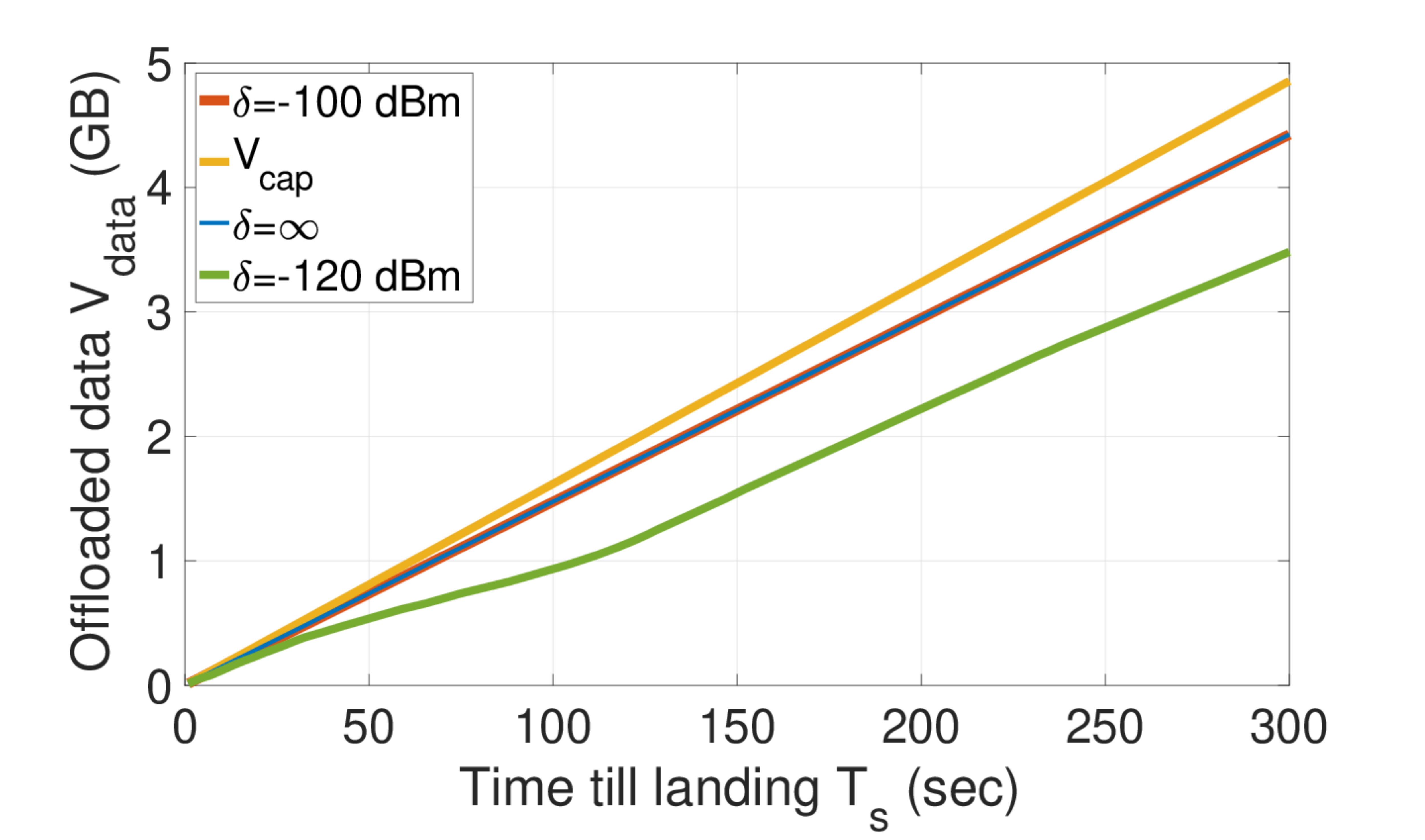} }
     \subfigure[Scenario 4]{
     \includegraphics[scale=0.15]{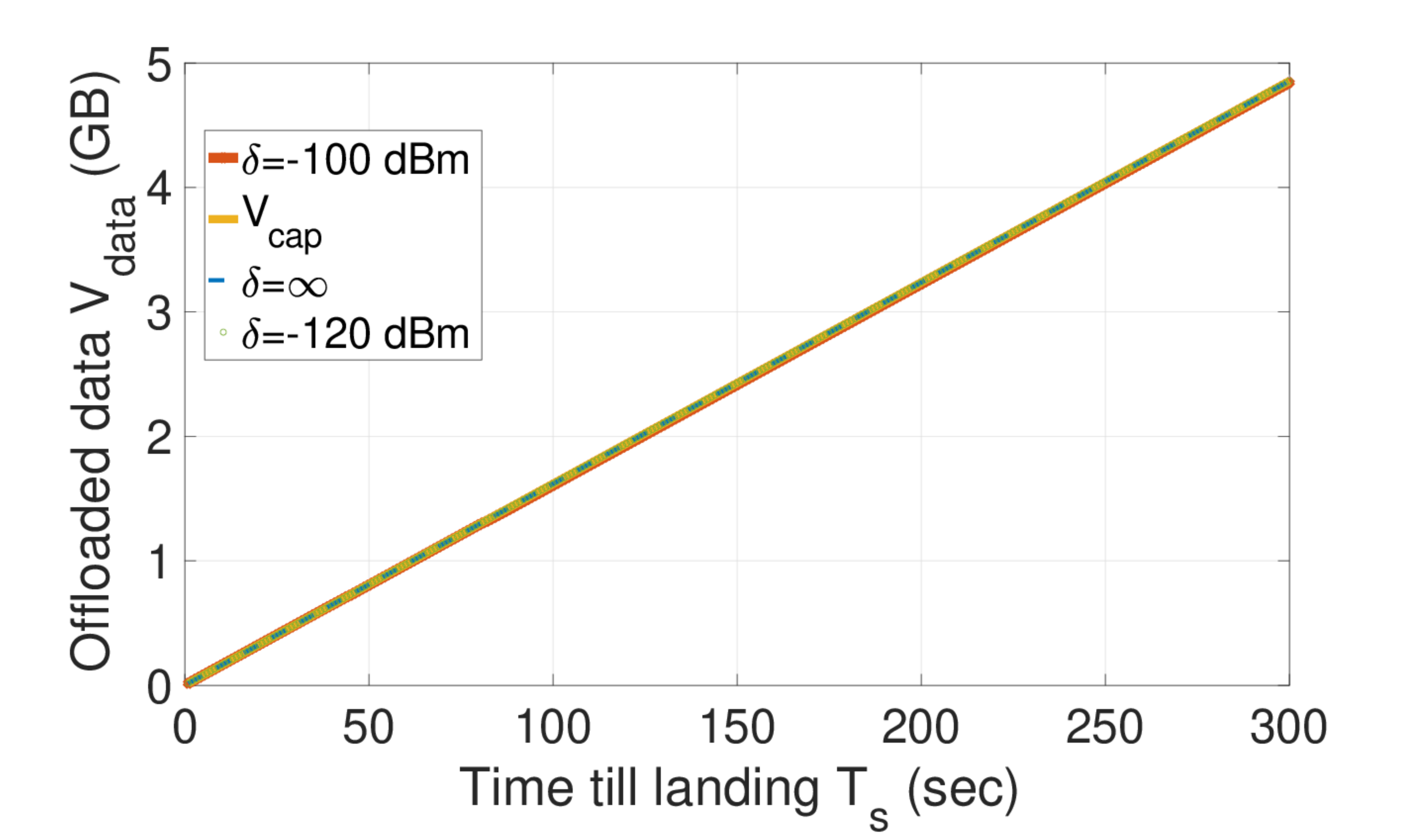} }
     \caption{Case~1: Impact of interference threshold $\delta$. The transmit power budget is set to $P_\text{max}=40$ W.}  \label{2ghz:impact:delta}
\end{figure}

In Fig.~\ref{2ghz:impact:delta}, we show the impact of interference threshold for three values of $\delta$: $\infty$ (no interference constraint), -100 dBm (the default interference threshold used in Fig.~\ref{figtoto}), and -120 dBm (a conservative threshold). In Scenario~1, the system is highly sensitive to the value of $\delta$ due to the weak directionality of directional antennas. For a conservative value of $\delta=-120$ dBm, $V_\text{data}$ drops to 50 MB (i.e., a significant loss w.r.t. $\delta=-100$ dBm). In Scenario~2, the impact is less severe although for $\delta=-120$ dBm there is still a significant loss w.r.t. $\delta=-100$ dBm. In Scenario~3, the A2G system performance is less impacted by the value of $\delta$  as the UPA installed at the plane dynamically adjusts its beam to steer the A2G interference away from the TBSs. This is even more pronounced for Scenario~4, where the performance is independent of the value of $\delta$.

\begin{figure}[!t]
         \centering
         \subfigure[Scenario 1]{
         \includegraphics[scale=0.15]{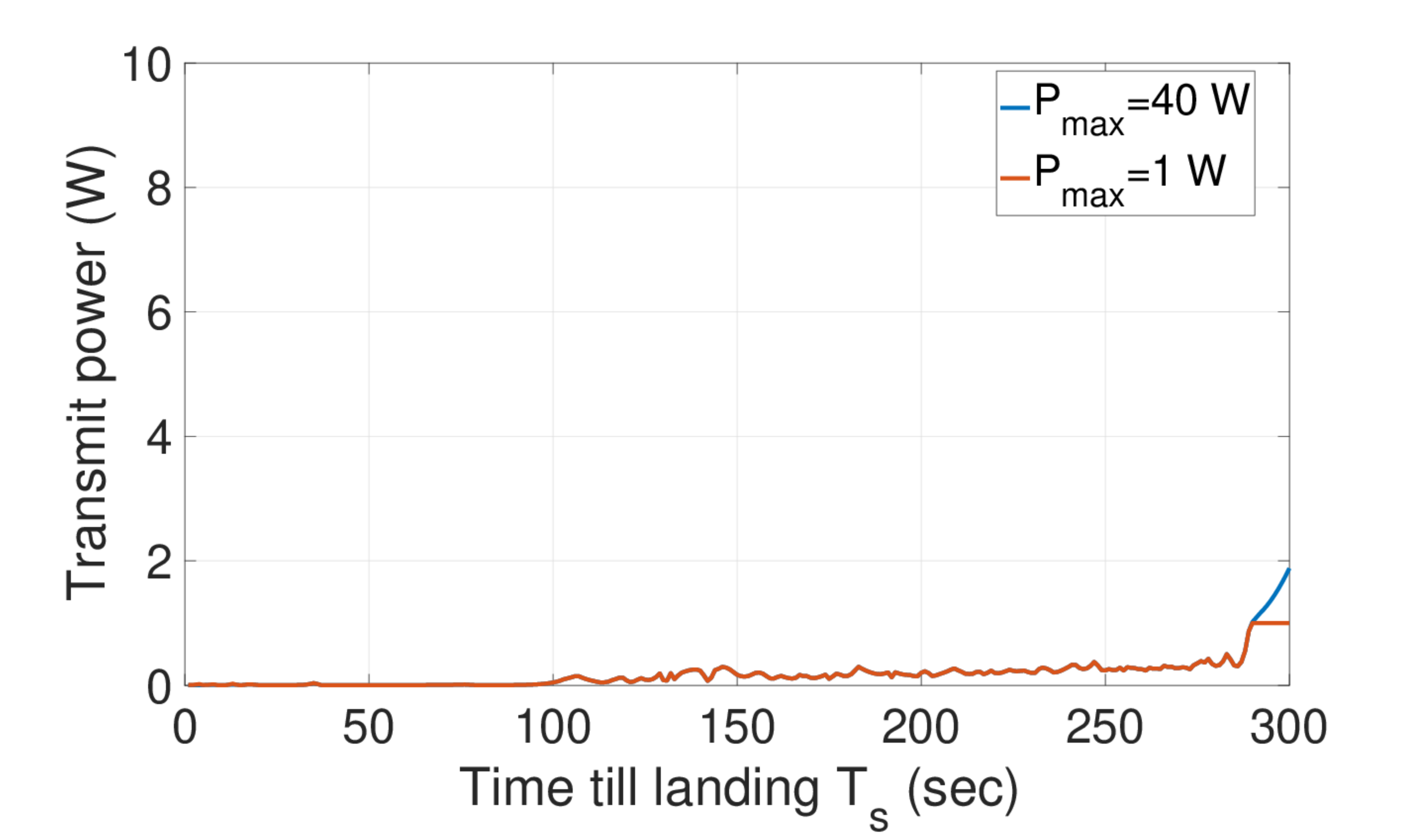} \label{2ghz:sec:sec:BF}}
         \subfigure[Scenario 2]{
         \includegraphics[scale=0.15]{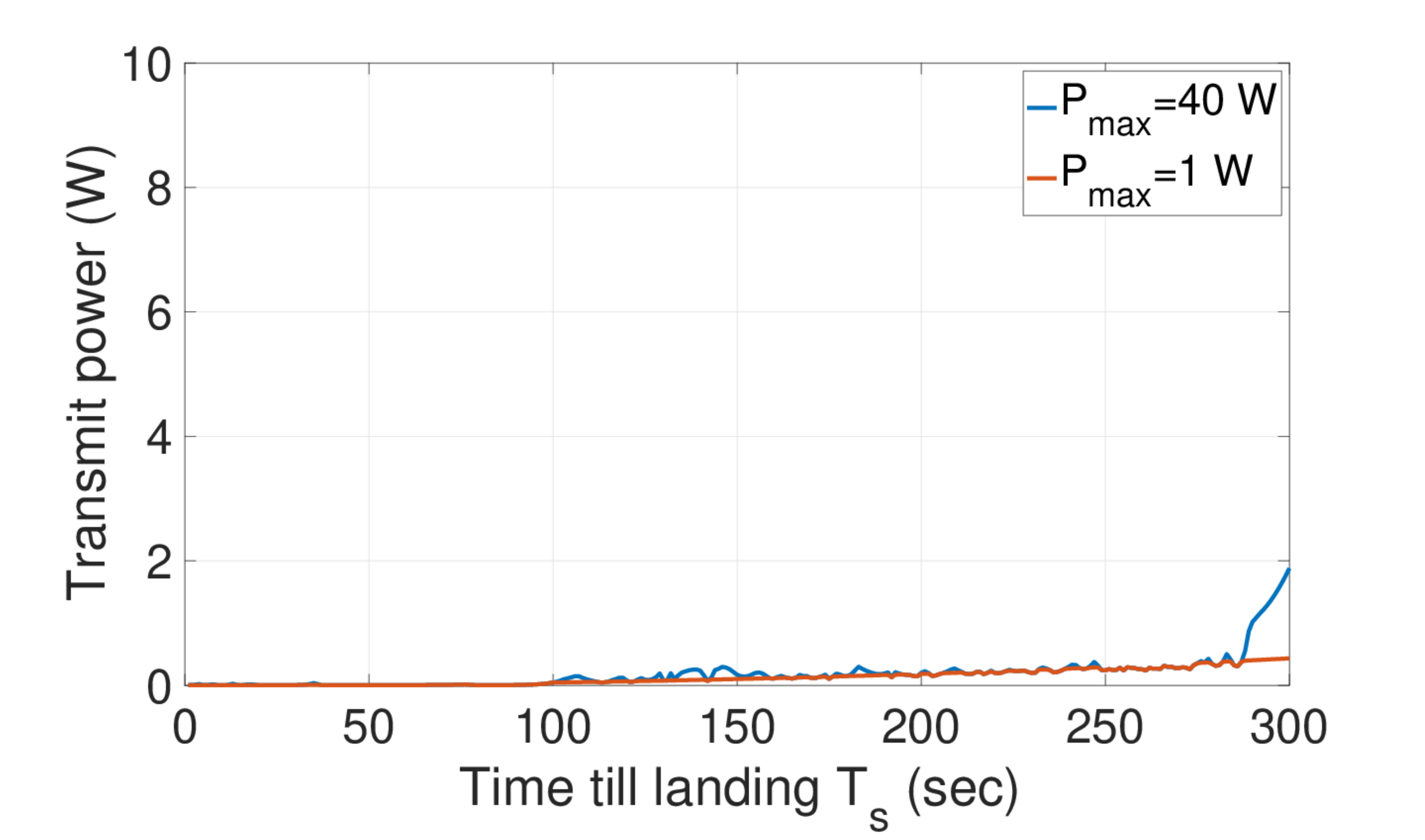} \label{2ghz:upa:sec:BF}}
         \subfigure[Scenario 3]{
         \includegraphics[scale=0.15]{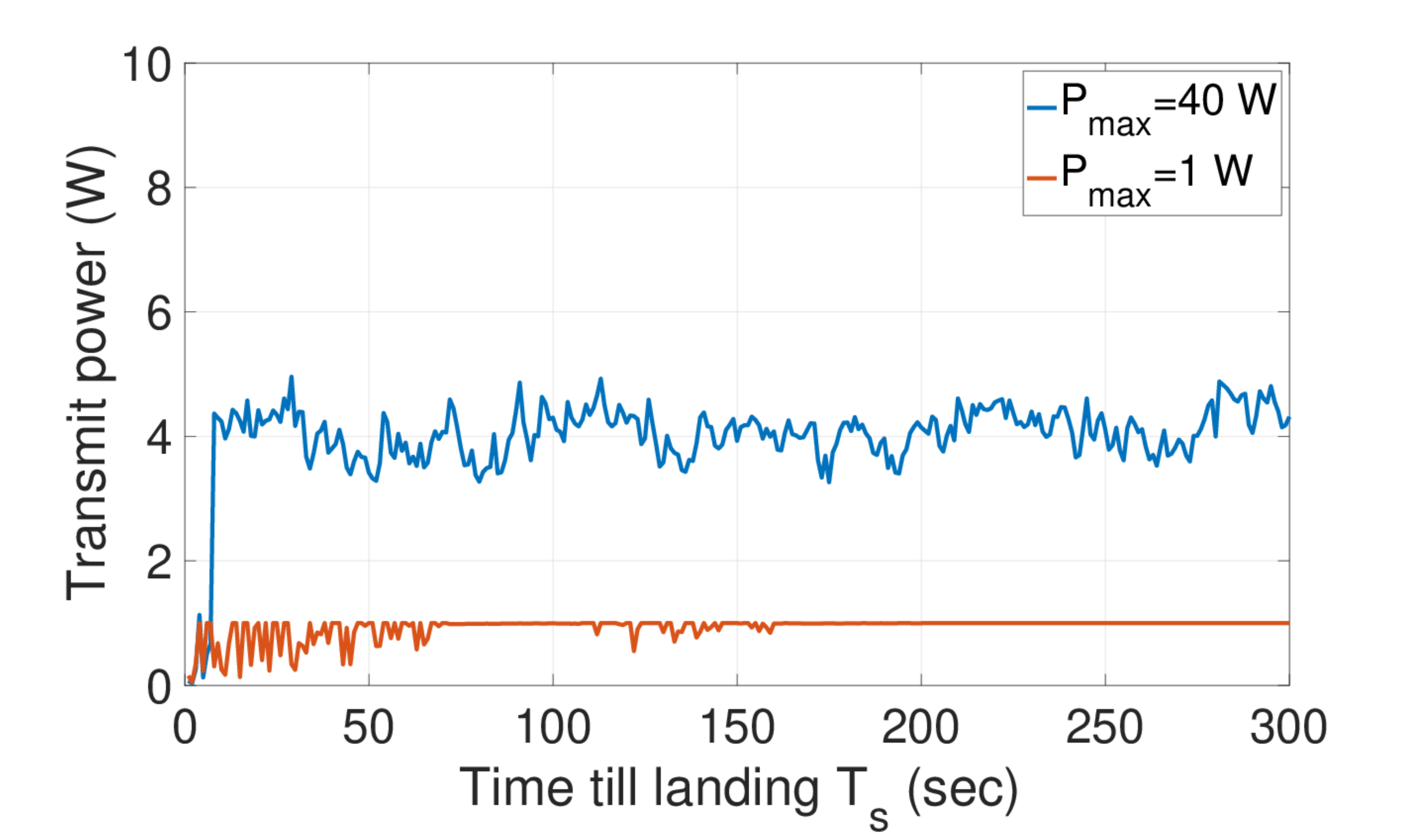} \label{2ghz:sec:upa:BF}}
         \subfigure[Scenario 4]{
         \includegraphics[scale=0.15]{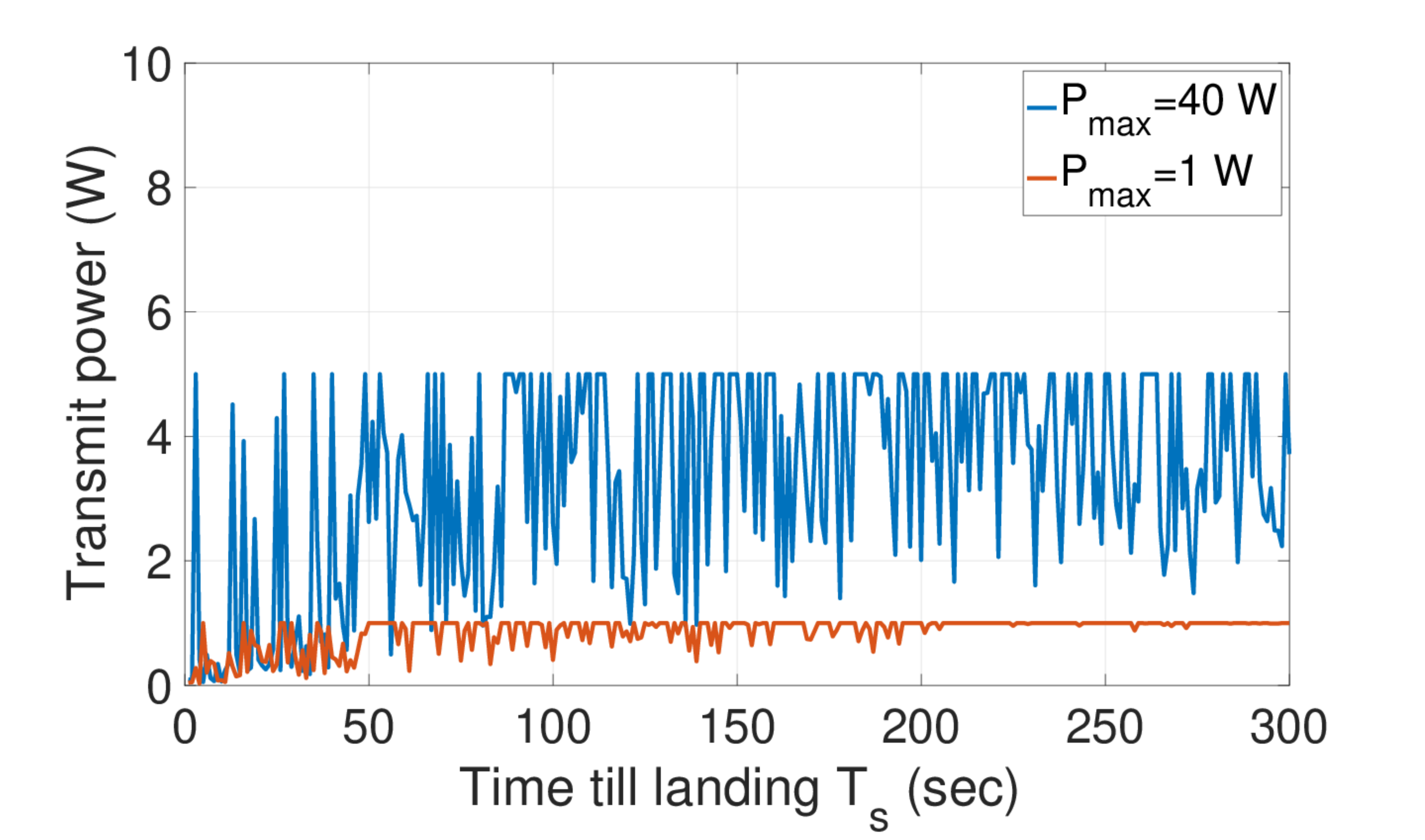} \label{2ghz:upa:upa:BF}}
         \caption{Case~1 --- Scenarios 1-4: Variation of transmitted power and transmitter/receiver antenna gain as a function of time.  
         }  \label{BF:gain}
    \end{figure}

We have observed that, for Case~1 in all scenarios and for the two power budgets, the plane transmits on the full bandwidth almost the entire time. This can be explained by the fact that in the microwave band due to the relatively small bandwidth and low path loss, the plane can still achieve a good SNR even with a small power budget (1 W).  
What this observation implies is that the system performance in all scenarios relies almost entirely on the adaptive tuning of transmit power or BF vectors at the plane and ABS. We can see this in Fig.~\ref{BF:gain} where we show the variation of transmit power of the plane during the descent for two values of $P_\text{max}$ for Scenarios~1-4. In all scenarios the transmit power is far less than 40 W throughout the descent thanks to the good channel conditions between the plane and ABS in the microwave band. 
In Scenarios~1 and~2 the variation of transmit power is slow for both values of $P_\text{max}$ while in Scenarios~3 and~4 it is rapid. This is because in the latter two scenarios, the plane has the degree of freedom to adjust its BF vector dynamically according to the position of the plane on the descent w.r.t. the ABS and TBSs. This also explains the robustness of the two scenarios to $\delta$. 

In conclusion, for Case~1:
\begin{itemize}
    \item The system is not power-limited with LTE-A and $P_\text{max}=1$~W is enough.
    \item We do not recommend using directional antennas at both the plane and ABS (Scenario~1) since then the system becomes highly interference-limited.
    \item The performance can almost reach $V_\text{cap}$ of LTE-A with a UPA at either the ABS or the plane. Using a UPA at the plane, however, makes the system more robust to $\delta$.
    \item  Using a UPA at both ends makes the system almost completely independent of $\delta$, eliminating the need for an exclusive microwave band for the A2G service. However, it does not bring much gain in the offloaded data with the existing LTE-A MCSs. If higher-order MCSs are introduced, the gain can be significant. 
    \item Dynamic tuning of transmit power and BF vectors at the ABS and/or the plane are crucial, while that of the number of subchannels to use is not.
\end{itemize}

\subsection{Case~2: Millimeter Band with Outdoor UPA-Based TBSs}
\begin{figure}
    \centering
    \includegraphics[scale=0.2]{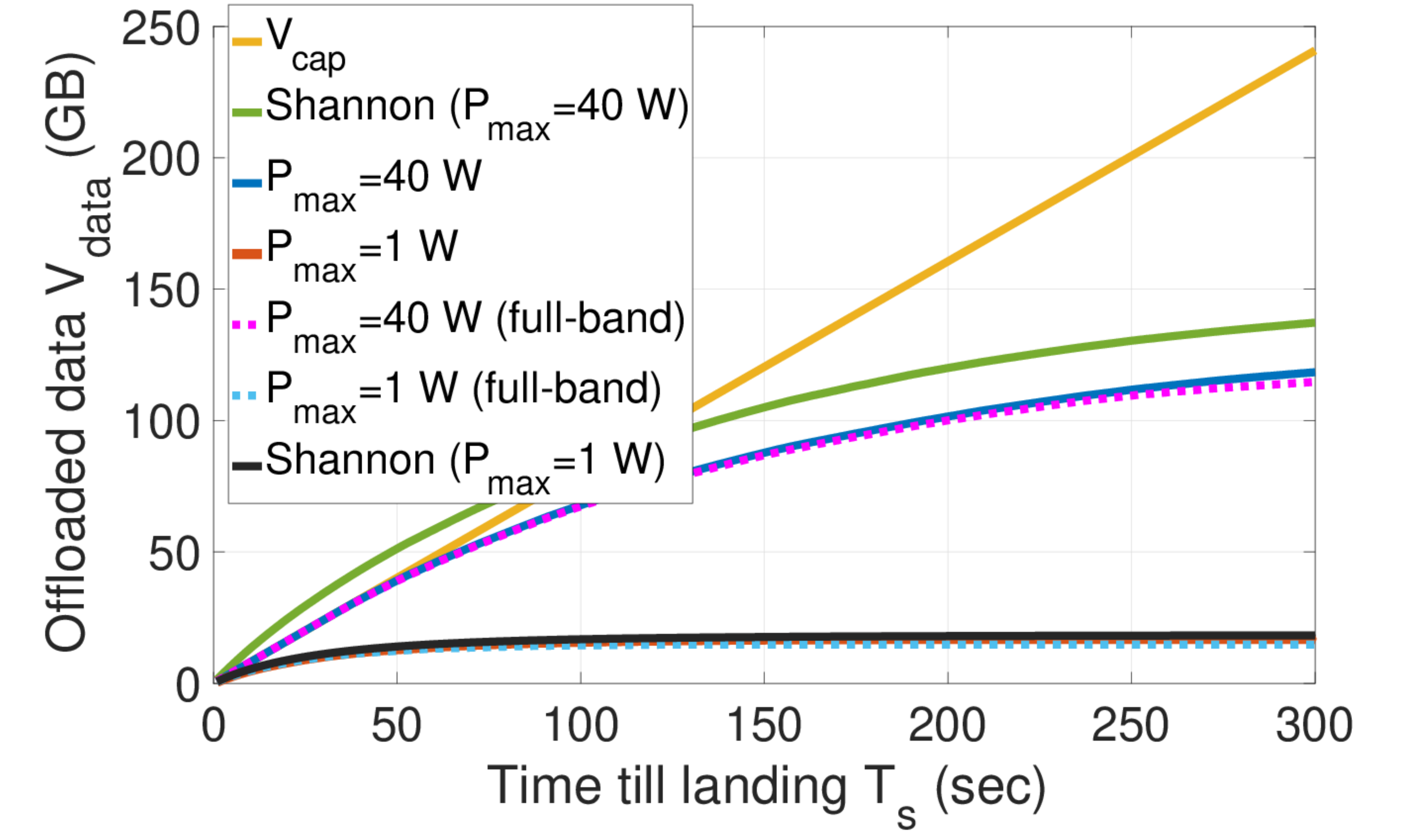}
    \caption{Case~2 --- Scenario 4: Impact of power on performance.}
    \label{28ghz:power}
\end{figure}

In this case, the system operates in the mmWave band. We assume every TBS is equipped with a 16-by-16 UPA\footnote{We use a relatively large array for the TBSs in the mmWave since we anticipate cellular operators would  do so to make up for the high path loss in this band in the uplink.}. 
We use a large 32-by-32 UPA at the ABS and a 5-by-5 UPA at the plane. 
In Fig.~\ref{28ghz:power}, we show the offloaded data volume $V_\text{data}$, $V_\text{cap}$, and the data volumes when using an ideal MCS function (Shannon) as a function of $T_s$ for two values of power budget $P_\text{max}$. 
We also show the offloaded data volume for a special case where the plane transmits on the full bandwidth throughout its descent. 
The operator can offload up to 120 GB of data using a power budget of 40 W in the last 300 seconds of descent with the LTE-A MCS function. 
The performance gain is 24 times that of Scenario 4 in Case~1 (on a bandwidth that is 50 times larger). 
With a transmit power budget of 1~W, the offloaded data is around 17 GB in the last 300~seconds of the descent which is just over 3 times that of Scenario 4 in Case~1. 
The results indicate that power budget has a significant impact on system performance. This is not surprising as the bandwidth is very large and the path loss in the 28-GHz band is high and has to be compensated by a large power budget per unit of bandwidth. 
An important observation is that $V_\text{data}$ reaches a plateau  (it is very visible for a power budget of 1~W). Most of the data is offloaded in the last 50~seconds (resp. 4~minutes) of the transmission period, for a power budget of 1~W (resp. 40~W), which is due to the small coverage in the 28 GHz band. The relative performance gain w.r.t. Shannon is marginal which suggests that adding higher-order MCSs in the mmWave band does not improve the performance as much as in the microwave band. 
The fact that all the curves, including the ones computed using Shannon's formula, are significantly below $V_\text{cap}$ indicates that the system is either interference-limited or power-limited (recall that $V_\text{cap}$ is computed assuming no SPC, PAPC, interference constraints and infinite power budget). Clearly for 1~W, the system is not at all MCS-limited (i.e., there is no gain to expect from higher order MCSs) while for 40~W, the performance could be marginally improved with higher order MCSs. To verify that the system is in fact power-limited (which is hinted by the significant difference in performance for the two power budgets), we tried different values of $\delta$,
i.e., $\delta=\infty$, -100, -120 dBm, for the two values of $P_\text{max}$. We observed that the performance remains almost unchanged even for the most conservative value of $\delta$, indicating that the gap we see in Fig.~\ref{28ghz:power} is not due to the interference constraints on the TBSs but rather due to the fact that the system is power-limited. 
Lastly, we note that allocating the whole band for transmission impacts the performance by 11\% and 18\% for $P_\text{max}$ of 40~W and 1~W respectively. The impact for $P_\text{max}=1$~W is more severe since a smaller power budget is spread over a large bandwidth. The result indicates that, unlike in Case~1 where full-bandwidth allocation to the plane was almost always optimal, in Case~2 fine-tuning the number of subchannels is crucial, especially for a low power budget. 

In conclusion, for Case~2:
\begin{itemize}
    \item The power budget has a significant impact on performance.
    \item The performance maintains a large gap from $V_\text{cap}$ of LTE-A, mainly because the system is highly power-limited.
    \item Adding higher-order MCSs to LTE-A can improve the performance particularly in the last minute of the descent if the power budget is high; otherwise, it does not. 
    \item Dynamic tuning of transmit power, BF vectors at the ABS and plane as well as the number of subchannels are all crucial. Similar to Scenario 4 of Case~1, there is no need for an exclusive band for the A2G service as the system is never interference-limited. 
\end{itemize}

\begin{table*}[!t]
\footnotesize
\caption{Simulation parameters }\label{params}
  \centering 
  \begin{tabular}{|c|c||c|c| }
    \hline
    Carrier frequencies $f_c$ & 2 \& 28 GHz &  Duration of a time slot $\Delta t$ & 1 ms  \\  
    \hline
    No. sub-carriers per channel & 12 &  Sub-carrier spacing & 15 kHz  \\
    \hline
    No. OFDM symbols per CTS & 14 & OFDM channel width $b$ & 180 kHz  
    \\ 
    \hline
    Bandwidth $ B(f_c)$ & 20 MHz \& 1 GHz & Max. no. Channels  $M$ & 112 \& 5556 \\ 
    \hline
    Interference threshold $\delta$ & $\infty$, -100, -120 dBm & No. ant. on the plane $N_\text{p}$ &  25 \\ 
    \hline
    No. ant. on the ABS $N_\text{A}$ &  1024& Runway length & 4 Km  \\  
    \hline
    Max. transmit power  $P_\text{max}$ & 1~W \& 40~W & Max. power per antenna element $P_\text{ant}$ & 0.2 W  \\ 
    \hline
    Absorption factor $L$ at $f_c=2$ GHz & 0.01 dB/Km & Absorption factor $L$ at $f_c=28$ GHz & 0.1 dB/Km  \\ 
    \hline
    Boresight gain of UPA elements & 8 dBi & Dir. boresight gain for the BSs & 17.7 dBi \\ 
    \hline
    Dir. boresight gain for the plane & 8 dBi & Tilt angle of plane's directional antenna & \ang{0} \\ 
    \hline
    Tilt angle of ABS Dir. antenna & \ang{3} & Tilt angle of UPAs & \ang{0} \\ 
    \hline
    BSs height & 30 m & Vertical velocity of the plane & -12.7 m/s\\
    \hline
    Pitch angle $\psi$ & \ang{3} & Noise PSD $\sigma^2$ & -174 dBm/Hz\\
    \hline    
    \specialcell{Path loss  $\beta_0(f_c,t)$} & \multicolumn{3}{ |c| }{ \specialcell{$ 32.5 + 20\log_{10}( \max(d_i(t),75) \times (f_c / 1000)) + L\times d_i(t)/1000  $ dB \\ ($d_i(t)$ is the distance between the plane and BS $i$ at $t$ in meter, $f_c$ in MHz) } }\\
    \hline
  \end{tabular}
\end{table*}

 
\section{Conclusion}\label{conclusion}
We considered an A2G system where a cellular operator facilitates offloading of a plane's maintenance data during its descent through an ABS at the airport. We developed a generic optimization framework that quantifies the system performance and includes two important ideas: (1) the A2G interference power on the legacy system remains less than a certain threshold $\delta$ and, (2) the transmit power, bandwidth, and BF vectors are dynamically tuned according to the distance between the plane and ABS. We tested our optimization framework for the microwave and mmWave bands, and various antenna settings at the plane and ABS. At the 2-GHz carrier (microwave) with a 20-MHz bandwidth, we can offload roughly 5~GB  of maintenance data over the last 5 minutes of descent with LTE-A MCSs. At the 28-GHz carrier (mmWave) with a 1-GHz bandwidth, we can roughly offload 120~GB of data over the same time interval using the same MCSs. It is best to use a UPA at the plane in the microwave both in terms of offloaded data volume and sensitivity to $\delta$. The impact of  power budget on performance in the mmWave band is significant while, in the microwave band, not much. In the microwave band, dynamic tuning of transmit power and BF vectors at the ABS and plane are crucial, while that of the number of subchannels to use is not. In the mmWave, dynamic tuning of all three variables is important. Adding higher-order MCSs to LTE-A can increase the offloaded data volume in the microwave band significantly, particularly in Scenario~4. In the mmWave, however, it improves the performance marginally. In both the microwave and mmWave, having a UPA at the plane provides a convenient control of the A2G interference through BF, eliminating the need for an exclusive spectrum for A2G. If directional antennas are used at the plane, a dedicated spectrum would significantly improve the performance. 

\appendices
\section{Proof of Lemma~\ref{lemma1}}\label{a1} 
\begin{proof}
We first calculate the noise power $n(t)$ at the ABS:
\allowdisplaybreaks
\begin{align*}
    &\mathbb{E}[|n(t)|^2] = \mathbb{E}[{n(t)}^H n(t)]
    = \mathbb{E}[\textbf{n}^H(t)\textbf{v}(t)\textbf{v}^H(t)\textbf{n}(t)]
    \stackrel{(a)}{=} \text{tr}\{\mathbb{E}[\textbf{n}^H(t)\textbf{v}(t)\textbf{v}^H(t)\textbf{n}(t)]\}\\
    &\stackrel{(b)}{=} \mathbb{E}[\text{tr}\{\textbf{n}^H(t)\textbf{v}(t)\textbf{v}^H(t)\textbf{n}(t)\}]
    \stackrel{(c)}{=} \mathbb{E}[\text{tr}\{\textbf{n}(t)\textbf{n}^H(t)\textbf{v}(t)\textbf{v}^H(t)\}]
    \stackrel{(d)}{=} \text{tr}\{\mathbb{E}[\textbf{n}(t)\textbf{n}^H(t)\textbf{v}(t)\textbf{v}^H(t)]\}\\
    &\stackrel{(e)}{=} \text{tr}\{\mathbb{E}[\textbf{n}(t)\textbf{n}^H(t)]\textbf{v}(t)\textbf{v}^H(t)\}
    \stackrel{(f)}{=} \text{tr}\{M(t)b\sigma^2 \textbf{I}_{N_\text{A}}\textbf{v}(t)\textbf{v}^H(t)\}\\
    &= M(t)b\sigma^2\text{tr}\{\textbf{v}(t)\textbf{v}^H(t)\}
    = M(t)b\sigma^2\textbf{v}^H(t)\textbf{v}(t) 
    = M(t)b\sigma^2||\textbf{v}(t)||^2,
\end{align*}
where, (a) holds since a scalar is equal to its trace, (b) follows from the linearity of expectation and trace operators (the order can be exchanged), (c) holds because of the cyclic property of the trace operator,
(d) follows from the linearity of the operations, (e) follows from the fact that $\textbf{v}(t)$ is not stochastic, and (f) follows from the distribution of noise vector $\textbf{n}(t)$.  
The SNR at the ABS, then, can be written as:
\begin{align}
    \gamma_0( t) &=  \frac{|\textbf{v}^H(t)\textbf{H}_0(t)\textbf{w}(t)|^2 \mathbb{E}[|s(t)|^2]}{\mathbb{E}[|n(t)|^2]} \notag 
    = \frac{|\textbf{v}^H(t)\textbf{H}_0(t)\textbf{w}(t)|^2}{Mb\sigma^2||\textbf{v}(t)||^2}\notag 
    = \frac{|\widetilde{\textbf{v}}^H(t)\textbf{H}_0(t)\textbf{w}(t)|^2}{M(t)b\sigma^2} \label{eq:gamma0},
\end{align}
where $\widetilde{\textbf{v}}(t)=\textbf{v}(t)/\|\textbf{v}(t)\|$ is a unit norm vector in the same direction as $\textbf{v}(t)$.
\end{proof}

\section{Proof of Lemma~\ref{lemma2}}\label{a2}
\begin{proof}
We prove the lemma by assuming otherwise and either reaching a contradiction or proving that choosing $\widetilde{\textbf{v}}_0$ as $\textbf{u}_\text{A}(t)/\sqrt{N_\text{A}}$ leads to the maximum possible objective value. 
If the claim is not true, then all of the optimal solutions for $\widetilde{\textbf{v}}(t)$ must have the form of $\widetilde{\textbf{v}}^{\text{opt}}(t) = a\textbf{u}_\text{A}(t) +b\bar{\textbf{u}}_\text{A}(t)$, where $a,b \in \mathbb{R}$, $b\not=0$, and $\bar{\textbf{u}}_\text{A}(t)$ is orthogonal to $\textbf{u}_\text{A}(t)$, i.e., $\bar{\textbf{u}}_\text{A}^H(t) \textbf{u}_\text{A}(t) = 0$. Furthermore, since $\textbf{u}^H_\text{A}(t)\textbf{u}_\text{A}(t) = N_\text{A}$, it is straightforward to see that $a<1/\sqrt{N_\text{A}}$ . Assuming that the plane applies the optimal beamforming vector $\textbf{w}^{\text{opt}}(t)$, we have 
\allowdisplaybreaks
\begin{align*}
    \gamma_0( t)  & = \frac{|(\widetilde{\textbf{v}}^\text{opt}(t))^H\textbf{H}_0(t)\textbf{w}^\text{opt}(t)|^2}{M(t) b \sigma^2}
    = \frac{ |(\widetilde{\textbf{v}}^\text{opt}(t))^H\textbf{u}_\text{A}(t)\textbf{u}_\text{P}^H(t)\textbf{w}^\text{opt}(t)|^2}{M(t) b \sigma^2}\\
    &= \frac{ |(a\textbf{u}_\text{A}(t) +b\bar{\textbf{u}}_\text{A}(t))^H\textbf{u}_\text{A}(t)\textbf{u}_\text{P}^H(t)\textbf{w}^\text{opt}(t)|^2}{M(t) b \sigma^2}
    = \frac{ |(a\overbrace{\textbf{u}_\text{A}^H(t)\textbf{u}_\text{A}(t)}^{N_\text{A}} +b\overbrace{\bar{\textbf{u}}_\text{A}^H(t)\textbf{u}_\text{A}(t)}^{0})\textbf{u}_\text{P}^H(t)\textbf{w}^\text{opt}(t)|^2}{M(t)b\sigma^2}\\
    &= \frac{ |aN_\text{A}\textbf{u}_\text{P}^H(t)\textbf{w}^\text{opt}(t)|^2}{ M(t)b\sigma^2}
    = \frac{ a^2N_\text{A}^2|\textbf{u}_\text{P}^H(t)\textbf{w}^\text{opt}(t)|^2}{M(t)b\sigma^2}  
    <\frac{ N_\text{A}|\textbf{u}_\text{P}^H(t)\textbf{w}^\text{opt}(t)|^2}{M(t) b\sigma^2}, 
\end{align*}
where the last inequality follows from $a<1/\sqrt{N_\text{A}}$. Let $\widetilde{\textbf{v}}'(t)= a'\textbf{u}_\text{A}(t)$ where $a'=1/N_\text{A}$. Suppose the plane still applies the previous beamforming vector, i.e., $\textbf{w}^\text{opt}(t)$, but the ABS applies the new beamforming vector $\widetilde{\textbf{v}}'(t)$ which is clearly feasible in the optimization Problem~$\boldsymbol{\Pi}$. It is straightforward to show that the SNR in this case is 
$\gamma_0'( t)=\frac{N_\text{A}|\textbf{u}_\text{P}^H(t)\textbf{w}^\text{opt}(t)|^2}{M(t)b\sigma^2}, $
which is greater than $\gamma_0(t)$, i.e., the achieved SNR when using beamforming vector $\widetilde{\textbf{v}}^\text{opt}(t)$ at the ABS. This either leads to a higher objective value in Problem~$\boldsymbol{\Pi}$ which is a contradiction, or does not change the objective value which means that the combination of $\widetilde{\textbf{v}}'(t)$ and $\textbf{w}^\text{opt}(t)$ is also optimal.
\end{proof}

\section{Proof of Lemma~\ref{lem:sce1} } \label{app:lem-sce1}
\begin{proof}
For a fixed $M(t) \in \mathcal{M}$, Problem~$\boldsymbol{\Pi}_\text{Sce1}$ reduces to 
\begin{equation} \label{p8}
    \begin{aligned} 
         \max_{ P(M(t),t)\leq P_\text{max} } &  M(t) b f\big( \frac{  P(M(t),t) \beta_0(f_\text{c},t) G_0^\text{P}(t)G^\text{A}(t)  }{M(t)b\sigma^2} \big)  \\ 
        \textrm{s.t.} \ \ 
        & \frac{P(M(t),t)G^\text{P}_{i^*}(t)G^\text{T}_{i^*}(t)}{M(t)}  \leq \delta,
    \end{aligned}
\end{equation} 
where $i^* \coloneqq \arg\max_{i \in \mathcal{I}} \{ G^\text{P}_i(t)G^\text{T}_i(t) \}$. (The index $i^*$ corresponds to the TBS(s) that is prone to the A2G interference the most.) The problem above is always feasible and the feasible solution must satisfy the last two constraints, i.e., $P(M(t),t)\leq \min \big( P_\text{max}, \frac{M(t)\delta}{ G^\text{P}_{i^*}(t) G^\text{T}_{i^*}(t) } \big)$. 
Since the objective function is increasing in $P(M(t),t )$, it is easy to see that the optimal solution to Problem~\eqref{p8} is $P^*(M(t),t )=\min \big( P_\text{max}, \frac{M(t)\delta}{ G^\text{P}_{i^*}(t) G^\text{T}_{i^*}(t) } \big)$. The optimal solution to Problem~$\boldsymbol{\Pi}_\text{Sce1}$, i.e., the pair $(M^*(t),P^*(t))$, can then be found by computing $P^*(M(t),t )$ for all $M(t)\in \mathcal{M}$ and selecting the pair that results in the highest objective value.
\end{proof}  

\bibliographystyle{IEEEtran} 
\bibliography{references}




 
\end{document}